\documentclass[a4paper,11pt]{article}
\pdfoutput=1

\usepackage{jheppub}
\usepackage{tikz}
\usepackage{verbatim}
\usetikzlibrary{shapes}
\usetikzlibrary{plotmarks}
\usepackage{tikz-3dplot}
\usepackage{footmisc}
\usepackage{ytableau}

\definecolor{light}{rgb}{0,1,1}

\title{Codimension-Two Defects and SYM on Orbifolds}
\author[1]{Roman Mauch}
\author[2]{and Lorenzo Ruggeri}
\affiliation[1]{Department of Physics and Astronomy, Uppsala Universitet, 752 37 Uppsala, Sweden}
\affiliation[2]{Yau Mathematical Sciences Center, Tsinghua University, Beijing, 100084, China}
\emailAdd{roman.mauch@physics.uu.se}
\emailAdd{ruggeri@mail.tsinghua.edu.cn}
\preprint{UUITP-08/25}

\abstract{We study $U(N)$ SYM theories on spaces with orbifold singularities via an equivalent description in terms of gauge theories on smooth manifolds with insertions of Gukov-Witten and twist defects. The combined effect of the defects is to render the fields multivalued with respect to rotations around the support of the defects. This motivates a relation with theories on branched covers, for which the multivaluedness has a geometric interpretation. We compute the partition function of the theory with defects on a patch and use it as a building block to compute partition functions on several closed spaces with conical singularities.}

\begin{document}

\maketitle
\flushbottom

\section{Introduction}
In the last few years, starting with \cite{Ferrero:2020laf,Ferrero:2020twa}, several works have investigated accelerating black hole solutions in supergravity with a near horizon geometry containing orbifold singularities. Many of these solutions involve a spindle $\mathbb{CP}^1_{\bsm p}$, a space topologically equivalent to a sphere but with conical singularities at the poles associated to deficit angles. Via AdS/CFT, quantities computed on the gravity side are reproduced by observables in a CFT. Recently, it has been shown that the entropy of AdS$_4$ black holes with near horizon geometry AdS$_2\times\mathbb{CP}^1_{\boldsymbol{p}}$ is reproduced by the large $N$ expansion of the spindle index \cite{Inglese:2023wky,Inglese:2023tyc,Colombo:2024mts}. Despite these exciting results, a clear understanding on how to define a gauge theory near an orbifold singularity is lacking. The main goals of this paper are to investigate the dynamics of a version of super Yang-Mills (SYM) theories near such singularities and to demonstrate an equivalence with certain SYM theories on branched covers.

Earlier works \cite{PeterBKronheimer:1990zmj,Douglas:1996sw,Kanno:2011fw} have studied the moduli space of instantons on orbifolds. In particular, the authors of \cite{Kanno:2011fw} considered an $\mathcal{N}=2$ $SU(\tilde{N})$ gauge theory on $(\mathbb{C}/\mathbb{Z}_{p})\times\mathbb{C}$ and, employing results in the mathematics literature \cite{Mehta,Biswas}, have shown how it can be equivalently described by the insertion on $\mathbb{C}^2$ of $\frac{1}{2}$-BPS codimension-two Gukov-Witten defects\footnote{Recently, a relation between theories on orbifolds and Gukov-Witten defects has also been suggested in \cite{Bomans:2024vii} from the gravity perspective. It would be interesting to compare their findings with those appearing in this work.} \cite{Gukov:2006jk,Gukov:2014gja}. These defects, specified by a partition $[n_I]$ of $\tilde{N}$ and by a choice of monodromy at $z_1=0$, break the gauge group to the so-called Levi subgroup $S[U(n_0)\times ...\times U(n_{p-1})]\subset U(\tilde{N})$ on the support of the defect $D$. As they impose a singular profile for the gauge connection on $D$, they can be thought of as higher codimensional analogues of 't Hooft lines.

In our work, we take $\tilde{N}=p N$, we select a specific partition such that $n_0=...=n_{p-1}=N$. The resulting $U(p N)$ gauge group is broken to its Levi subgroup $\mathbb{L}=U(N)^p$. This particular choice admits a $\mathbb{Z}_p$-symmetry cyclically permuting the fields in each $U(N)$-factor. Thus, the theory admits twist defects \cite{Dixon:1986qv,Cardy:2007mb,Calabrese:2009qy,Giveon:2015cgs}, which we insert on $D$. Finally, we integrate out the fields in $U(p N)/\mathbb{L}$ by giving a large vev to the scalar. We call the resulting theory the refined $U(N)$ orbifold theory. The insertion of twist defects enforces multivaluedness of the fields on $\mathbb{C}^2$ in the path integral: when a field in the $k^\text{th}$ $U(N)$-factor encircles the support of the twist defect, it is identified with a field in the $(k+1)^\text{st}$ $U(N)$-factor. On $(\mathbb{C}/\mathbb{Z}_p)\times\mathbb{C}$ this translates into a field returning to its original value only after encircling the singularity $p$ times. Via a field redefinition we can achieve that $\mathbb{Z}_p$ acts diagonally, enforcing twisted boundary conditions in each $U(N)$-factor. We generalize this construction\footnote{Adding a free $S^1$ fiber along which the defect extend, we also consider odd-dimensional setups.} to 2d for an $\mathcal{N}=(2,2)$ vector multiplet and in 4d for two intersecting GW defects \cite{Koh:2008kt,Gomis:2016ljm,Jeong:2020uxz}.

We use the refined orbifold theory to draw a comparison between the behaviour of SQFTs on orbifolds and branched covers. The $\mathbb{Z}_p$-symmetry, which makes the field multivalued on the orbifold, is related to a similar $\mathbb{Z}_p$-symmetry on the branched cover of $\mathbb{C}^2$. For the latter, the fields in each sheet are cyclically identified, on the branch cut, with those on the following sheet \cite{Casini:2009sr,Nishioka:2013haa,Nishioka:2016guu}. It is known that observables on two-dimensional branched covers are reproduced by observables in a theory obtained by gauging this $\mathbb{Z}_p$-symmetry \cite{Cardy:2007mb,Calabrese:2009qy,Giveon:2015cgs}, giving a theory on a smooth manifold with multivalued fields. We argue that this theory, on the level of partition functions, is equivalent to the refined orbifold theory. A similar equivalence has been pointed out in \cite{Hamidi:1986vh,Dixon:1986qv,Bershadsky:1986fv,Atick:1987kd} for strings propagating on orbifolds or branched covers of Riemann surfaces.

The equivalence above explains an observation made in \cite{Mauch:2024uyt}. There, it has been shown that partition functions on certain compact spaces with conical singularities with deficit angles can be obtained dimensionally reducing from compact spaces with conical singularities with surplus angles. In this paper we use the refined orbifold partition function on $\mathbb{C}/\mathbb{Z}_{p}$ and $\mathbb{C}/\mathbb{Z}_{p_1}\times\mathbb{C}/\mathbb{Z}_{p_2}$ as building blocks to compute more examples of equivariant SYM theories on compact spaces with conical singularities. We discuss how to glue the building blocks in cases where the resulting space is a symplectic toric orbifold \cite{Lerman97,Martelli:2023oqk}, adopting the factorization properties from the case of smooth manifolds \cite{Pestun:2007rz,Pasquetti:2011fj,Benini:2012ui,Beem:2012mb,Festuccia:2018rew,Festuccia:2019akm}. This way, we can compute the partition function of the refined orbifold theory on orbifolded spheres and weighted projective spaces. As expected, this reproduces results on branched covers of spheres \cite{Nishioka:2016guu} and on weighted projective spaces \cite{Inglese:2023wky,Martelli:2023oqk}. Another example is presented in detail in \cite{Ruggeri:2025zon} for spaces whose topology is the product of two spindles.

The outline is as follows. In \autoref{sec.2} we introduce our definition of refined orbifold theory with the insertion of Gukov-Witten and twist defects, and show its relation to spaces surplus angles. We also explain its relation with the orbifold theory in \cite{Kanno:2011fw}. Then, in \autoref{sec.3}, we compute the partition function for the refined orbifold theory on $\mathbb{C}/\mathbb{Z}_{p}$, $\mathbb{C}/\mathbb{Z}_{p_1}\times\mathbb{C}/\mathbb{Z}_{p_2}$ and in odd dimensions, including instanton contributions in 4d and 5d. We use these building blocks, in \autoref{sec.4}, to compute partition functions on compact spaces with conical singularities. Here, with our approach, we reproduce results already known in the literature while also filling several gaps.

\paragraph{Notation.}
Throughout this article, $\boldsymbol p=(p_1,\dots,p_r)$ always denotes an array of integers $p_i$ such that $\gcd(p_i,p_j)=1$ (unless specified otherwise), and $|\boldsymbol p|=p_1\cdot\ldots\cdot p_r$. $\widehat{\mathbb{C}}^{2}_{\boldsymbol p}$ denotes the branched cover of $\mathbb{C}^2$ with branch locus $\{z_1=0\}\cup\{z_2=0\}$, where the two $\mathbb{C}$-planes have respective branch indices $p_1,p_2$ and intersect orthogonally at the origin. $\widehat{\mathbb{C}}_p$ denotes a branched cover of $\mathbb{C}$ with locus at the origin and branch index $p$.

\section{Codimension Two Defects and Conical Singularities}\label{sec.2}

In this section we consider the $\mathcal{N}=(2,2)$ vector multiplet on $\mathbb{C}$ and $\mathcal{N}=2$ vector multiplet on $\mathbb{C}^2$ in the presence of both a GW defect and a twist defect. After giving a specific vev to the scalar in the vector multiplet, we argue that this theory equivalently describes the respective vector multiplet on a branched cover $\widehat{\mathbb{C}}_{\boldsymbol p}$, respectively $\widehat{\mathbb{C}}^2_{\boldsymbol p}$, and an orbifold $\mathbb{C}/\mathbb{Z}_{p}$, respectively $\mathbb{C}/\mathbb{Z}_{p_1}\times\mathbb{C}/\mathbb{Z}_{p_2}$.

\subsection{Gukov-Witten Defects}
\label{sub.GW}

Gukov-Witten (GW) defects were first introduced in \cite{Gukov:2006jk} for 4d $\mathcal{N}=4$ SYM as $\frac{1}{2}$-BPS operators supported on a codimension two submanifold $D$ and can be viewed as higher codimension analogues of 't Hooft line operators. They were later generalized to $\frac{1}{2}$-BPS operators in $\mathcal{N}=2$ \cite{Gukov:2007ck,Gaiotto:2009fs}, to $\frac{1}{4}$-BPS operators \cite{Koh:2008kt,Gomis:2016ljm,Jeong:2020uxz} and to codimension-two operators in dimensions ranging from 2d to 6d \cite{Drukker:2008jm,Kapustin:2012iw,Drukker:2012sr,Bullimore:2014upa,Bullimore:2014awa,Hosomichi:2015pia}. 

In fact, there is an equivalence between gauge theories in the presence of GW defects and on orbifolds---based on the equivalence between parabolic bundles and bundles on orbifolds \cite{Biswas}.

\paragraph{Two Dimensions.}
Consider the 2d $\mathcal{N}=(2,2)$ vector multiplet on $\mathbb{C}$ with $G=U(M)$ and let us place a GW defect at the origin. The defect is classified by an element $\xi$ in the Cartan subalgebra of $U(M)$:
\begin{equation}\label{eq.defect2d}
    \xi=\diag(\underbrace{\xi_{(0)},\dots,\xi_{(0)}}_{M_0\text{ times}},\underbrace{\xi_{(1)},\dots,\xi_{(1)}}_{M_1\text{ times}},\dots,\underbrace{\xi_{(p-1)},\dots,\xi_{(p-1)}}_{M_{p-1}\text{ times}}).
\end{equation} 
Close to the defect, we can choose polar coordinates $(r,\theta)$ on the normal plane in which the gauge field takes the form $A\sim \xi\,\dd\theta$. For the field strength we obtain the following singular profile:
\begin{equation}
    F\sim\xi\,\delta(r)\,\dd r\wedge\dd\theta.
\end{equation}
This can be completed into a $\frac{1}{2}$-BPS configuration by choosing a similar profile for the auxiliary field \cite{Hosomichi:2015pia}. Fields charged under the gauge group acquire a holonomy\footnote{Both the scalar and the fermions in the vector multiplet have charge 1 under the Cartan subalgebra of $U(M)$. For a generic value of the charge $q$, the phase induced by the GW defect would depend explicitly on its value.} 
\begin{equation}
    \exp\left(\ii\int_\gamma A\right)=\e{2\pi\ii\xi}
\end{equation} 
when moving along a small loop $\gamma$ around the origin.

In order to preserve the defect, gauge transformations on its support must be restricted to take values in the subgroup of $U(M)$ that commutes with $\xi$, called the Levi subgroup $\mathbb{L}$. For the choice in \eqref{eq.defect2d} we have 
\begin{equation}\label{eq.2dLevi}
    \mathbb{L}=\prod_{k=0}^{p-1}U(M_k).
\end{equation}
The theory only depends on the partition $[M_0,\dots,M_{p-1}]$ of $M$, but not its ordering. 

Later on, we want to make contact with a $U(N)$-gauge theory on the branched cover $\widehat{\mathbb{C}}_p$. This requires the specific choice of
\begin{equation}
    M_0=\dots=M_{p-1}=:N,\qquad \xi_{(k)}=\frac{k}{p},
\end{equation}
for which the Levi subgroup is given by $\mathbb{L}=U(N)^p\subset U(pN)$. As mentioned previously, the theory is independent of the ordering of the different blocks in \eqref{eq.defect2d}. In other words, the theory has a symmetry determined by outer automorphisms of $\mathbb{L}$, and since gauge transformations are restricted to values in $\mathbb{L}=U(N)^p$ on the defect, this is a global $\mathbb{S}_p$-symmetry. In the following, we are interested only in the $\mathbb{Z}_p$-subgroup generated by
\begin{equation}
    \label{eq.T2d}
    T=\begin{pmatrix}
        0 & 1_{N} & 0 & 0 & \cdots & 0\\
        0 & 0 & 1_{N} & 0 & \cdots & 0\\
        \vdots &  &  &  & \ddots & \vdots\\
        1_{N} & 0 & 0 & 0 & \cdots & 0 
    \end{pmatrix}\in U(pN)
\end{equation}
which acts on the field content of the vector multiplet in the adjoint.

\paragraph{Four Dimensions.}
Consider the 4d $\mathcal{N}=2$ vector multiplet on $\mathbb{C}^2$ with $G=U(M)$. We want to consider two GW defects placed on the codimension two surfaces $D_j$ defined by $z_j=0$ ($j=1,2$) \cite{Koh:2008kt,Gomis:2016ljm,Jeong:2020uxz}. In analogy to the 2d case, we want to make contact with a $U(N)$-gauge theory on the branched cover $\widehat{\mathbb{C}}^2_{(p_1,p_2)}$ later on. Therefore, we only consider defects classified by Cartan elements
\begin{subequations}
    \label{eq.defect4d}
    \begin{align}
        \xi&=\diag(\underbrace{\xi_{(0)},\dots,\xi_{(0)}}_{p_2N},\underbrace{\xi_{(1)},\dots,\xi_{(1)}}_{p_2N},\dots,\underbrace{\xi_{(p_1-1)},\dots,\xi_{(p_1-1)}}_{p_2N}),\\
        \zeta&=\diag(\underbrace{\zeta_{(0)},\dots,\zeta_{(p_2-1)}}_{p_1N},\underbrace{\zeta_{(0)},\dots,\zeta_{(p_2-1)}}_{p_1N},\dots,\underbrace{\zeta_{(0)},\dots,\zeta_{(p_2-1)}}_{p_1N}),
    \end{align}
\end{subequations}
with $\xi_{(k)}=\frac{k}{p_1}$ ($k=0,\dots,p_1-1$) and $\zeta_{(k)}=\frac{l}{p_2}$ ($l=0,\dots,p_2-1$), and $\gcd(p_1,p_2)=1$.

Close to the intersection of the defects, the gauge field takes the form $A\sim \xi\, \dd\theta_1+\zeta\,\dd\theta_2$ (where $z_j=r_j\e{\ii\theta_j}$). The corresponding field strength is
\begin{equation}\label{eq.F4d}
    F\sim\xi\,\delta(r_1)\,\dd r_1\wedge\dd\theta_1+\zeta\,\delta(r_2)\,\dd r_2\wedge\dd\theta_2.
\end{equation}
This field configuration can be completed into a $\frac{1}{4}$-BPS multiplet which preserves a $U(1)^2$-subgroup of the isometry group $SO(4)$ \cite{Gomis:2016ljm,Jeong:2020uxz}. Fields charged under the gauge group acquire a holonomy
\begin{equation}
    \exp\left(\ii\int_{\gamma_1}A\right)=\e{2\pi\ii\xi},\qquad\exp\left(\ii\int_{\gamma_2} A\right)=\e{2\pi\ii\zeta}
\end{equation} 
when moving along a small loop $\gamma_j$ around $D_j$. Notice that $\pi_1(\mathbb{C}^2-(D_1\cup D_2))=\mathbb{Z}\times\mathbb{Z}$ is abelian, in agreement with the commuting holonomies around $\gamma_1,\gamma_2$.

On $D_1$ the gauge group is broken to $\mathbb{L}_1=U(p_2N)^{p_1}$ while on $D_2$ it is broken to $\mathbb{L}_2=U(p_1N)^{p_2}$. On the intersection $D_1\cap D_2=\{0\}$ the Levi subgroup is\footnote{This is true if $\gcd(p_1,p_2)=1$. For $\gcd(p_1,p_2)=p>1$, then $\mathbb{L}_{12}=U(pN)^{\frac{p_1p_2}{p}}$.} $\mathbb{L}_{12}=U(N)^{p_1p_2}$. In analogy to the 2d case, the theory is independent\footnote{Note that this is not necessarily true if the blocks in \eqref{eq.defect4d} are of different size, as different orderings can yield inequivalent parabolic subgroups \cite{Kanno:2011fw}.} of the ordering of the various blocks in \eqref{eq.defect4d} and thus possesses a global symmetry. The subgroup relevant for us is $\mathbb{Z}_{p_1}\times\mathbb{Z}_{p_2}$, generated by
\begin{equation}
    \label{eq.BC.defects}
    T_1=\begin{pmatrix}
        & 1_{p_2N} &  &  &  & \\
        &  & 1_{p_2N} &  &  & \\
        &  &  &  & \ddots & \\
        &  &  &  &  & 1_{p_2N}\\
        1_{p_2N} &  &  &  &  &  
    \end{pmatrix},\qquad
    T_2=\begin{pmatrix}
        P &  &  &  & \\
        & P &  &  & \\
        & & & \ddots & \\
        &  &  &  & P
    \end{pmatrix},
\end{equation}
where $P$ is the $p_2N$-dimensional square matrix obtained from $T_1$ by replacing $1_{p_2N}$ by $1_N$. The generators act on the field content of the vector multiplet in the adjoint.

\subsection{Twist Defects}
\label{sub.twist}

Whenever a quantum field theory has a global internal symmetry, that is a symmetry that acts in the same way everywhere in space and which does not change the positions of fields, we can introduce twist defects \cite{Calabrese:2009qy}. They typically appear in two-dimensional QFTs as local operators such that, when encircling them, the fields of the theory change by a symmetry transformation \cite{Dixon:1986qv,Bershadsky:1986fv,Cardy:2007mb,Calabrese:2009qy}. Here, like GW defects, we take a twist defect to be supported on codimension two submanifolds $D$. These defects can be thought of as GW defects for a global symmetry\footnote{In 3d these are often denoted \emph{global vortex loops} \cite{Kapustin:2012iw,Drukker:2012sr}.}, and the conditions under which these codimension two defects, associated to either global or gauge symmetries, are topological have been studied recently \cite{Heidenreich:2021xpr}. In particular, the presence of charged scalars and fermions in the vector multiplets we are considering makes the defects non-topological. A similar construction for a non-supersymmetric pure gauge theory would make at least a subset of them topological. However, even if our setup they are not topological, they can be inserted and employed to prescribe a singular behaviour for the (background) gauge field.

\paragraph{Two Dimensions.}
In the previous subsection we saw the existence of a global $\mathbb{Z}_p$-symmetry for the 2d theory with GW defect, generated by \eqref{eq.T2d}. Let us therefore introduce a twist defect $\mathcal{T}$ at the origin, such that it implements the following transformations on the field content $\Phi$ of the $U(pN)$-vector multiplet when encircling $\mathcal{T}$:
\begin{equation}\label{eq.2dtwisthol}
    \mathcal{T}[D]:\;\Phi\mapsto \Ad_T\Phi.
\end{equation} 
Note that, for the naive definition of $\mathcal{T}$ above, the fermionic fields might pick up an additional minus sign when rotating around $\mathcal{T}$, spoiling supersymmetry. However, one should always be able to make $\mathcal{T}$ supersymmetric by dressing it with some additional operators (e.g. involving $R$-symmetry transformations \cite{Giveon:2015cgs}), such that both, bosons and fermions satisfy the periodicity condition $\mathcal{T}^p(\Phi)=\Phi$. In the following, when writing the $\mathcal{T}$-action, we only specify the naive part\footnote{This naive part describes the action of $\mathcal{T}$ on the scalar field (which has vanishing $R$-charge and is unaffected by the dressing). It turns out that this is enough to determine the effect of $\mathcal{T}$ on the partition function.} for brevity. The transformation in \eqref{eq.2dtwisthol} can be understood as arising from the holonomy of a background gauge field taking values in the cyclic group generated by \eqref{eq.T2d}, explaining their interpretation as GW defects for a global symmetry.

\paragraph{Four Dimensions.}
In the four-dimensional theory with GW defects as in \autoref{sub.GW}, we have a global $\mathbb{Z}_{p_1}\times\mathbb{Z}_{p_2}$-symmetry, generated by $T_1,T_2$ in \eqref{eq.BC.defects}. We can thus define two twist defects $\mathcal{T}_1,\mathcal{T}_2$ supported, respectively, on $D_1,D_2$ which act on the fields as
\begin{align}
    \label{eq.4d.twist}
    \mathcal{T}_1[D_1]:\;\Phi\mapsto\Ad_{T_1}\Phi,\qquad
    \mathcal{T}_2[D_2]:\;\Phi\mapsto\Ad_{T_2}\Phi.
\end{align}
Similarly to the 2d case, also here the fermions might be anti-periodic under $\mathcal{T}_j^{p_j}$ ($j=1,2$), which again can be remedied by dressing the naive twist defect with additional operators, such that $\mathcal{T}_j^{p_j}\cdot\Phi=\Phi$. Moreover, as in 2d, their interpretation as global GW defects arises by considering an holonomy taking values in \eqref{eq.BC.defects}.

\subsection{A Refined Orbifold Theory}
\label{sub.orbi}
Consider again the $U(\bsm p N)$-vector multiplet on $\mathbb{C}^r$ in the presence of the GW defect as introduced above. This theory turns out to be equivalent \cite{Kanno:2011fw,Hosomichi:2015pia,Biswas} to the $U(\bsm pN)$-vector multiplet on the orbifold $\mathbb{C}^r_{\bsm p}$. We want to reformulate our 2d and 4d theory with GW and twist defects on such an orbifold. Moreover, both on the orbifold and the GW theories, we will introduce a specific vev for the scalar which breaks the gauge group to the Levi subgroup everywhere.

\paragraph{Two Dimensions.}
For the 2d theory with GW defect as described in \autoref{sub.GW}, there exists an equivalent description as a $\mathcal{N}=(2,2)$ vector multiplet on an orbifold $\mathbb{C}/\mathbb{Z}_p$ \cite{Biswas,Hosomichi:2015pia}. The orbifold action on $z\in\mathbb{C}$ and $a\in\mathfrak{u}(pN)$ is given by
\begin{equation}
    \label{eq.2d.orbi}
    z\mapsto\omega z,\qquad a\mapsto\Ad_{\exp(2\pi\ii \xi)} a,
\end{equation}
where $\omega=\e{2\pi\ii/p}$ and $\xi$ as in \eqref{eq.defect2d}. Due to the equivalence of the two theories, the global $\mathbb{Z}_p$-symmetry discussed in \autoref{sub.GW} persists in the orbifold theory. 
 
Next, we introduce a vev for the scalar $X$ in the vector multiplet of the form
\begin{equation}\label{eq.higgs2d}
    X^0=\diag(\underbrace{h_0,\dots,h_0}_{N\text{ times}},\dots,\underbrace{h_{p-1},\dots,h_{p-1}}_{N\text{ times}}),
\end{equation}
where we take\footnote{The limit $h_i\to\infty$ should be understood as a convenient shorthand: any non-zero vev already breaks the gauge group to the Levi subgroup and renders the off-diagonal modes massive, with mass proportional to $|h_i|$. At finite vev these modes are still present at intermediate energies, but they decouple in the IR. Hence, the IR fixed point is insensitive to the scale of the vev. A similar argument applies to the 4d setup discussed below. See section 3.1 in \cite{Nishioka:2016guu} for a related setup where the Higgsing with infinite vev is employed.} $h_i\to\infty$. Hence, the gauge group is broken to $\mathbb{L}=U(N)^p$ everywhere, and the infinitely massive fields do not contribute to the computation. The Higgsing is compatible with the orbifold action \eqref{eq.2d.orbi} and, in fact, on the remaining light fields the action is trivial. Note that we really consider $X_0$ up to actions of the Weyl group of $U(pN)$, and the global $\mathbb{Z}_p$-symmetry generated by \eqref{eq.T2d} descends to the Higgsed theory. Moreover, the theory with GW defect can be equivalently described as coupling the $U(pN)$-gauge theory to a codimension-two theory supported on the insertion locus of the defect \cite{Gukov:2006jk}. The Higgsing above ``freezes'' the degrees of freedom of this lower-dimensional theory, so we do not have to account for them in the partition function. We will give an interpretation for this phenomenon when considering the one-loop determinant in \autoref{sec.3.1}.

For this $U(N)^p$-gauge theory on $\mathbb{C}/\mathbb{Z}_p$, let us introduce a twist defect at the origin of $\mathbb{C}/\mathbb{Z}_p$. Denoting by $\Phi_{k}$ the field whose only non-trivial entry is in the $k^\text{th}$ factor of $\mathfrak{u}(N)^p$, the action \eqref{eq.T2d} of the twist defect, e.g., when encircled by the scalar field $\phi_k$, is given by
\begin{equation}
    \label{eq.2d.twist}
    \mathcal{T}[D]:\;\phi_{k}\mapsto\phi_{k+1}.
\end{equation}
Namely, in the presence of $\mathcal{T}$, the field $\Phi$ only returns to its original form after encircling $\mathcal{T}$ a number of $p$ times, hence becomes multi-valued. In fact, we can interpret $\mathcal{T}$ itself as an orbifold action on top of \eqref{eq.2d.orbi}. We call this theory---the $\mathcal{N}=(2,2)$ vector multiplet on $\mathbb{C}/\mathbb{Z}_p$ with actions \eqref{eq.2d.orbi} and \eqref{eq.2d.twist}---the $U(N)$ \textit{refined orbifold theory}, in order to distinguish it from the one with only the orbifold action \eqref{eq.2d.orbi}. 

It is particularly interesting to study the effect of twist defects on the allowed gauge transformations. Considering, for simplicity the abelian case ($N=1$), before the insertion of a twist defect a gauge transformation $g$ along the polar angle is given by
\begin{equation}
    g(\theta)=\left(e^{\ii\frac{m_0}{p}\theta},\dots,e^{\ii\frac{m_{p-1}}{p}\theta}\right),\qquad m_k\in p\mathbb{Z}.
\end{equation}
Instead, in the refined orbifold theory, the boundary conditions \eqref{eq.2d.twist} have the effect of imposing
\begin{equation}\label{eq.quantization}
    m_0=\dots=m_{p-1}\equiv m,\qquad m\in\mathbb{Z}.
\end{equation}
Hence, also gauge transformation with fractional charge are allowed. This will have a crucial relevance when considering the flux quantization condition on spaces with non-trivial two-cycles in \autoref{sec.4}. This discussion can be easily generalized to the non-abelian case.

Once we localise this theory, when considering fluctuations around the locus, it is useful to perform a reparametrisation of the fields by
\begin{equation}
    \label{eq.2d.redef}
    \tilde\phi_k=\sum_{l=0}^{p-1}\e{2\pi\ii kl/p}\phi_l
\end{equation}
(and similarly for the remaining field content). This is simply a global $SU(pN)$-transformation of $\phi$. The action of the twist defect on $\tilde\phi_k$ is given by
\begin{equation}
    \label{eq.2d.Tdiag}
    \mathcal{T}[D]:\;\tilde\phi_k\mapsto\e{2\pi\ii k/p}\tilde\phi_k,
\end{equation}
i.e. $\mathcal{T}$ acts diagonally, and the fluctuations for different $U(N)$-factors decouple\footnote{While we can perform the redefinition \eqref{eq.2d.redef} already before localisation, note that the $U(N)$-factors are still coupled in the full theory due to interactions between the factors.}. We will comment on the effect of \eqref{eq.2d.Tdiag} on the partition function in the next section.

\paragraph{Four Dimensions.}
Our definition of the refined orbifold proceeds analogously to the 2d case. First, we observe that the 4d theory on $\mathbb{C}^2$ with GW defect \eqref{eq.defect4d} is equivalently described by the theory on an orbifold $\mathbb{C}/\mathbb{Z}_{p_1}\times\mathbb{C}/\mathbb{Z}_{p_2}$, with the $\mathbb{Z}_{p_1}\times\mathbb{Z}_{p_2}$-action on $(z,w)\in\mathbb{C}^2$ and $a\in\mathfrak{u}(p_1p_2N)$ given by
\begin{subequations}
    \label{eq.4d.orb}
    \begin{align}
        (z,w)\mapsto(\omega z, w),\qquad &a\mapsto\Ad_{\exp(2\pi\ii\xi)}a,\label{eq.4d.orbT}\\
        (z,w)\mapsto(z,\eta w),\qquad &a\mapsto\Ad_{\exp(2\pi\ii\zeta)}a.
    \end{align}
\end{subequations}
Here, $\omega=\e{2\pi\ii/{p_1}}$, $\eta=\e{2\pi\ii/{p_2}}$ and $\xi$, $\zeta$ as in \eqref{eq.defect4d}. The equivalence was explained for a single defect $\xi$ in \cite{Kanno:2011fw} and more generally is based on the equivalence between parabolic bundles on the smooth space and bundles on the orbifold \cite{Biswas}. 

Next, we perform a Higgsing with scalar vev
\begin{equation}\label{eq.higgs4d}
    X^0=\diag(\underbrace{h_{0,0},\dots,h_{0,0}}_{N\text{ times}},\underbrace{h_{0,1},\dots,h_{0,1}}_{N\text{ times}},\dots,\underbrace{h_{p_1-1,p_2-1},\dots,h_{p_1-1,p_2-1}}_{N\text{ times}}),
\end{equation}
with $h_{i,j}\to\infty$. This breaks the gauge group to $U(N)^{p_1p_2}$ and only the light fields contribute to the computation. As in the 2d case, the Higgsing is compatible with the orbifold action and the Higgsed theory still has a global $\mathbb{Z}_{p_1}\times \mathbb{Z}_{p_2}$-symmetry. Note that the orbifold action \eqref{eq.4d.orb} on the light fields is trivial. Also in this case, the theory with GW defects can be viewed as a $U(p_1p_2N)$-gauge theory coupled to codimension-two and codimension-four theories living on $D_1$, $D_2$ and $D_1\cap D_2$. Higgsing again ``freezes'' these lower-dimensional degrees of freedom (see \autoref{sec.3.1}).

Now let us introduce the two twist defects \eqref{eq.4d.twist} for this global symmetry into the $U(N)^{p_1p_2}$-gauge theory on $\mathbb{C}/\mathbb{Z}_{p_1}\times\mathbb{C}/\mathbb{Z}_{p_2}$. Notating by $\phi_{k,l}$ the scalar field whose non-trivial entry is in the $(k\!\cdot\!l)^\text{th}$ factor of $\mathfrak{u}(N)^{p_1p_2}$, the action of the twist defect, when encircled, is given by
\begin{equation}
    \label{eq.4d.twist.explicit}
    \mathcal{T}_1[D_1]:\;\phi_{k,l}\mapsto\phi_{k+1,l},\qquad\mathcal{T}_2[D_2]:\;\phi_{k,l}\mapsto\phi_{k,l+1}.
\end{equation}
In fact, we can interpret $\mathcal{T}_1[D_1],\mathcal{T}_2[D_2]$ as orbifold actions on top of \eqref{eq.4d.orb}. As in 2d, we call this theory the $U(N)$ refined orbifold theory. The same quantization for charges appearing in a gauge transformation \eqref{eq.quantization} applies in the four-dimensional setup.

For the purpose of localisation, it is again useful to consider the following field redefinition:
\begin{equation}
    \label{eq.4d.redef}
    \tilde{\phi}_{k,l}=\sum_{r=0}^{p_1}\sum_{s=0}^{p_2}\e{2\pi\ii kr/p_1}\e{2\pi\ii ls/p_2}\phi_{r,s},
\end{equation}
(and similarly for the remaining field content) which is simply a global $SU(p_1p_2N)$ transformation. The action of the twist defects on $\tilde\phi_{k,l}$ is given by
\begin{equation}
    \label{eq.4d.Tdiag}
    \mathcal{T}_1[D_1]:\;\tilde\phi_{k,l}\mapsto \e{2\pi\ii k/p_1}\tilde\phi_{k,l},\qquad\mathcal{T}_2[D_2]:\;\tilde\phi_{k,l}\mapsto\e{2\pi\ii l/p_2}\tilde\phi_{k,l}.
\end{equation}
Hence, the theory of quadratic fluctuations around the localisation locus decouples for the different $U(N)$-factors upon the field redefinition \eqref{eq.4d.redef}.

\paragraph{Adding a Free Circle.}
The 2d (4d) refined orbifold theory can be extended in a natural way to a 3d $\mathcal{N}=2$ (5d $\mathcal{N}=1$) vector multiplet by considering an extension of spacetime $\mathbb{C}/\mathbb{Z}_p$ ($\mathbb{C}/\mathbb{Z}_{p_1}\times\mathbb{C}/\mathbb{Z}_{p_2}$) by an $S^1$. Namely, we consider the respective fields on a trivial $U(1)$-orbibundle over $\mathbb{C}/\mathbb{Z}_p$ ($\mathbb{C}/\mathbb{Z}_{p_1}\times\mathbb{C}/\mathbb{Z}_{p_2}$). The orbifold singularity in 3d (5d) is supported on a circle, and the theory now has a $T^2$-isometry ($T^3$-isometry) with respect to which we can localise. Due to the trivial extension, the only additional contribution in the partition function is a product over representations of the new $U(1)$ (see next section).

\subsection{Comparison with Branched Cover}\label{sub.branched}

We anticipate that the refined orbifold theory defined above is, in fact, equivalent to a $U(N)$-vector multiplet on a branched cover. Let us therefore briefly review the latter and draw a comparison between the two theories. We do this for the 4d case; the 2d case is completely analogous.

Let us consider the 4d $\mathcal{N}=2$ vector multiplet with gauge group $U(N)$ on $\widehat{\mathbb{C}}^2_{\bsm p}$. This space is a $p_1 p_2$-sheeted branched cover of $\mathbb{C}^2$ for which $p_1$ sheets collapse on $D_1$ and $p_2$ on $D_2$. Its monodromy $m:\pi_1(\mathbb{C}^2-(D_1\cup D_2))\rightarrow \mathbb{S}_{p_1p_2}$ is defined by mapping the two generators $a,b$ as follows:
\begin{equation}
    \label{eq.4d.branching}
    a\mapsto\sigma_1,\qquad b\mapsto\sigma_2.
\end{equation}
Here, $\sigma_1$ acts on $\{1,2,\dots,p_1p_2\}$ by shifting the $i^\text{th}$ block ($i=1,\dots,p_1$) of numbers $\{j,j+1,\dots,j+p_2\}$ with $j=0\mmod p_2$ to the $(i+1)^\text{st}$ block; $\sigma_2$ acts within the $i^\text{th}$ block, giving $\{j+p_2,j,\dots,j+p_2-1\}$.
The fields on $\widehat{\mathbb{C}}^2_{\bsm p}$ have to satisfy appropriate boundary conditions at the branch locus $D_1\cup D_2$. 

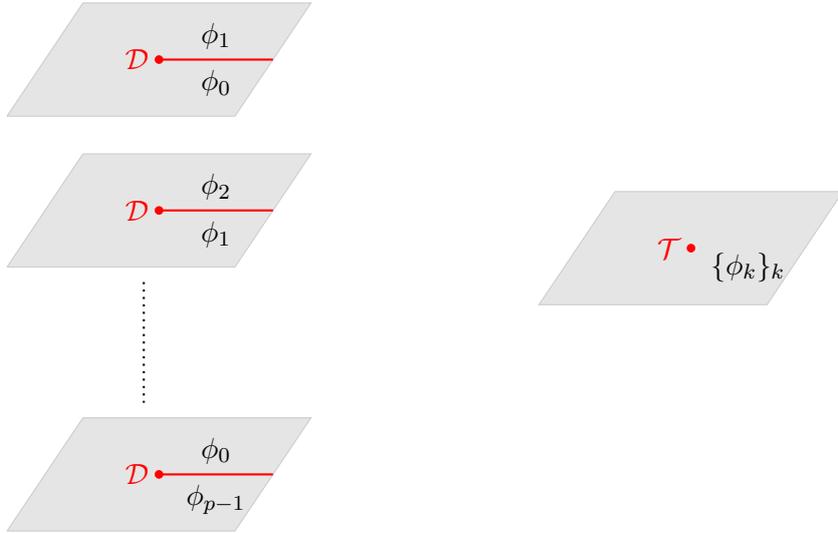
\begin{figure}[h]
    \centering
    \begin{tikzpicture}[scale=1]
                    \draw[fill=gray,opacity=.2] (-2, 4) -- (1, 4) -- (2, 5.5) --(-1, 5.5) -- (-2, 4);
                    \draw[thick,red] (0,4.75)--(1.5,4.75)  node[black,midway, below]{$\phi_{0}$} node[black,midway, above]{$\phi_{1}$};
                    \draw[red,fill=red] (0,4.75) circle [radius=0.05] node[left]{$\mathcal{D}$};
                    \draw[fill=gray,opacity=.2] (-2, 2) -- (1, 2) -- (2, 3.5) --(-1, 3.5) -- (-2, 2);
                    \draw[thick,red] (0,2.75)--(1.5,2.75)  node[black,midway, below]{$\phi_{1}$} node[black,midway, above]{$\phi_{2}$};
                    \draw[fill=red,red] (0,2.75) circle [radius=0.05] node[left]{$\mathcal{D}$};
                    \draw[dotted, thick] (-0.2, 1.8) -- (-0.2, 0.2);
                    \draw[fill=gray,opacity=.2] (-2, -1.5) -- (1, -1.5) -- (2, 0) --(-1, 0) -- (-2, -1.5);
                    \draw[thick,red] (0,-0.75)--(1.5,-0.75)  node[black,midway, below]{$\phi_{p-1}$} node[black,midway, above]{$\phi_{0}$};
                    \draw[fill=red,red] (0,-0.75) circle [radius=0.05] node[left]{$\mathcal{D}$};
        \begin{scope}[shift={(7,0)}]
                    \draw[fill=gray,opacity=.2] (-2,1.5) -- (1,1.5) -- (2,3) --(-1,3) -- cycle;
                    \draw[fill=red,red] (0,2.25) circle [radius=0.05] node[left]{$\mathcal{T}$};
                    \draw (0.75,2) node{$\{\phi_{k}\}_k$};
        \end{scope}
    \end{tikzpicture}
    \caption{Left-hand side: $p$ copies of the theory on $\mathbb{C}$, with a codimension one defect $\mathcal{D}$ implementing the branching structure. Right-hand side: theory for the collection $\{\phi_k\}_k$ of all fields on a single sheet $\mathbb{C}$, with a twist defect $\mathcal{T}$ implementing multivaluedness.}
    \label{fig.nishioka}
\end{figure}

Equivalently, we can describe this theory by $p_1 p_2$ copies of the $U(N)$-vector multiplet on $\mathbb{C}^2$, with the branching structure implemented by codimension one defects $\mathcal{D}_i$ supported on the product of $D_i$ with a half-line, acting\footnote{Note that we are sweeping several details under the rug here, in particular concerning SUSY, in favour of a simple exposition. Rigorous definitions of the theory on branched covers can usually be given in terms of a resolved space (see e.g. \cite{Nishioka:2013haa}).} as 
\begin{equation}
    \label{eq.4d.BC}
    \mathcal{D}_1[D_1\times \mathbb{R}]:\;\phi_{k,l}\mapsto\phi_{k+1,l},\qquad \mathcal{D}_2[\mathbb{R}\times D_2]:\;\phi_{k,l}\mapsto\phi_{k,l+1}.
\end{equation}
on fields $\phi_{k,l}$ in the $k\!\cdot\! l$-copy; see \autoref{fig.nishioka} for the simpler 2d setup. 

This is reminiscent of the action of twist defects in \eqref{eq.4d.twist.explicit}. In fact, it is easy to see that the setup above is equivalent to a collection of (interacting) $\mathfrak{u}(N)$-valued fields $\{\phi_{k,l}\}$ defined on a single sheet $\mathbb{C}^2$, which we can collect into a $\mathfrak{u}(N)^{p_1p_2}$-valued field $\phi$ that is multivalued when encircling $D_1\cup D_2$, according to \eqref{eq.4d.BC} \cite{Casini:2005rm}. The multivaluedness of $\phi$ can be traded for the existence of precisely the twist defects \eqref{eq.4d.twist.explicit} on $\mathbb{C}^2$ \cite{Cardy:2007mb,Calabrese:2009qy,Giveon:2015cgs}. In fact, the steps above can be interpreted in terms of a discrete gauging of the theory of $p_1p_2$ copies: this theory has a global $\mathbb{S}_{p_1p_2}$-symmetry that permutes the different copies. Then the theory on a single sheet $\mathbb{C}^2$ with twist defects $\mathcal{T}_{1,2}$ is obtained by gauging the $\mathbb{Z}_{p_1}\times\mathbb{Z}_{p_2}$-subgroup that acts on the fields as in \eqref{eq.4d.BC} \cite{Giveon:2015cgs}.

Finally, this theory of $\mathfrak{u}(N)^{p_1p_2}$-valued fields on $\mathbb{C}^2$ with twist defects $\mathcal{T}_{1,2}$---in terms of the twisted boundary conditions of the fields---can equivalently be viewed as the theory of $\mathfrak{u}(N)^{p_1p_2}$-valued fields on the orbifold $\mathbb{C}/\mathbb{Z}_{p_1}\times\mathbb{C}/\mathbb{Z}_{p_2}$, with orbifold action as in \eqref{eq.4d.twist.explicit}. Hence, we expect, e.g., the partition function of the vector multiplet on the branched cover $\widehat{\mathbb{C}}^2_{\bsm p}$ with \eqref{eq.4d.branching} to agree with the one of the refined orbifold theory (and this is indeed what we observe in \autoref{sec.3}). 

Note that in \cite{Hamidi:1986vh,Atick:1987kd} a similar observation has been made also for strings propagating on orbifolds \cite{Dixon:1986qv} and on branched covers \cite{Bershadsky:1986fv} of Riemann surfaces.

\section{Partition Functions}
\label{sec.3}

In this section we compute the partition function for our 2d and 4d refined orbifold theories. We will find agreement with the respective theory on the branched cover, as discussed in the previous section. Finally we will comment on how these results can be extended to 3d and 5d by considering the theory on trivial $S^1$-orbibundles. The results of this section will serve as building blocks for theories on compact orbifolds, which we discuss in \autoref{sec.4}.

Note that all defects are supported on subspaces that are invariant under the Cartan torus of the isometry group. This is paramount for equivariance of the theory.

\subsection{Two Dimensions}\label{sec.3.1}

We can now analyse the partition function of the refined orbifold theory. Introducing the large vev for the scalar, but without the twist defect, the orbifold action on the fields is trivial, and the partition function can be obtained via localisation, e.g., from the result for vortex defects in \cite{Hosomichi:2015pia}. For the one-loop determinant we have
\begin{equation}\label{eq.prova}
    \prod_{\alpha\in\Delta_{\mathbb{L}}}\,\prod_{n=0}^\infty(n\epsilon+\ii\alpha(a)),
\end{equation}
where $\Delta_{\mathbb{L}}$ is the set of roots of $\mathbb{L}$, $\epsilon$ the equivariance parameter for the $U(1)$-action, and $a$ the Coulomb branch parameter.

As mentioned below \eqref{eq.higgs2d}, we now explain why the Higgsing causes the degrees of freedom living on the defect to decouple from the theory. Without the Higgsing, the one-loop determinant would receive contributions also from roots valued in $\Delta_{U(pN)}/\Delta_{\mathbb{L}}$. Unlike fields valued in $\mathbb{L}$, these contributions would be non-trivially affected by the presence of the GW defect \eqref{eq.2d.orbi} and the ratio $Z^\text{1-loop}(\xi)/Z^\text{1-loop}(\xi=0)$ exactly gives the contribution from fluctuations on the codimension-two defect \cite{Nawata:2014nca,Hosomichi:2021gxe}. Since this ratio is 1 for the Higgsed theory we conclude that all degrees of freedom on the defect are integrated out in the Higgsed theory. The same argument applies in higher dimensions and for instanton contributions \cite{Nawata:2014nca}.

We now go back to the Higgsed theory and the one-loop determinant in \eqref{eq.prova}. In the presence of the twist defect, the Coulomb branch parameter is restricted to take the same value in all $\mathfrak{u}(N)$-factors, effectively introducing a delta-function $\prod_{k=0}^{p-1}\delta(a_k-a_{k+1})$ (where $a_{p}=a_0$). 
The one-loop determinant is obtained by considering fluctuations around the localisation locus to quadratic order. Here, the redefinition \eqref{eq.2d.redef} actually decouples the different $U(N)$-factors in the theory, and the one-loop determinant is obtained by a product of the $p$ factors. Each of those can be computed essentially by counting local holomorphic functions at the origin, weighted by their charge under the $U(1)$-action from equivariance. For the $k^\text{th}$ $U(N)$-factor, given the periodicity constraint \eqref{eq.2d.Tdiag}, a basis of holomorphic functions has charges $\epsilon(n+\frac{k}{p})$. Thus, the one-loop determinant for the refined orbifold reads
\begin{equation}
    \label{eq.2d.Z.fact}
    Z_{p}(a;\epsilon)=\prod_{k=0}^{p-1}\,\prod_{\alpha\in\Delta_{U(N)}}\,\prod_{n=0}^\infty\left(n\epsilon+\frac{k}{p}\epsilon+\ii\alpha(a)\right),
\end{equation}  
where we now understand $a$ as the Coulomb branch parameter for $U(N)$. Note that this can be rewritten as
\begin{equation}
    \label{eq.2d.Z.unfact}
    Z_{p}(a;\epsilon)=\prod_{\alpha\in\Delta_{U(N)}}\,\prod_{n=0}^\infty\left(n\frac{\epsilon}{p}+\ii\alpha(a)\right),
\end{equation}
which is precisely the one-loop determinant of the $\mathcal{N}=(2,2)$ vector multiplet with gauge group $U(N)$ on the branched cover $\widehat{\mathbb{C}}_{\bsm p}$ (cf. \cite{Nishioka:2013haa}). Hence, we see that the partition functions for the refined orbifold and branched cover theories agree.

\subsection{Four Dimensions}

The partition function of the 4d refined orbifold theory can similarly be obtained by localisation. The orbifold action, without the twist defect, is trivial on the fields. For the one-loop determinant for the Higgsed theory we have
\begin{equation}
    \prod_{\alpha\in\Delta_{\mathbb{L}_{12}}}\Upsilon(\ii\alpha(a)|\epsilon_1,\epsilon_2),
\end{equation}
where $\epsilon_1,\epsilon_2$ denote the equivariance parameters for the $T^2$-action and $\Upsilon(x|y,z)$ the upsilon function (see \autoref{app.upsilon}).

In analogy to the 2d case, once twist defects are present, we effectively introduce a delta-function for the Coulomb branch parameters $a_{k,l}$ of the respective $\mathfrak{u}(N)$-factors into the integrand, $\prod_{k=0}^{p_1-1}\prod_{l=0}^{p_2-1}\delta(a_{k,l}-a_{k+1,l})\delta(a_{k,l}-a_{k,l+1})$ (where $a_{p_1,l}=a_{0,l}$ and $a_{k,p_2}=a_{k,0}$). Again, in complete analogy to the 2d case, after performing the field redefinition \eqref{eq.4d.redef} for the fluctuations, fields in different $\mathfrak{u}(N)$-factors decouple. Due to the periodicity constraints \eqref{eq.4d.Tdiag}, a basis of local holomorphic functions for the $(k,l)$-factor has charges $\epsilon_1(n_1+\frac{k}{p_1})+\epsilon_2(n_2+\frac{l}{p_2})$ and the one-loop determinant for the refined orbifold reads
\begin{equation}
    \label{eq.4d.Z.fact}
    Z_{\bsm p}(a;\epsilon_1,\epsilon_2)=\prod_{k=0}^{p_1-1}\;\prod_{l=0}^{p_2-1}\,\prod_{\alpha\in\Delta_{U(N)}}\Upsilon\Big(\ii\alpha(a)+\epsilon_1\frac{k}{p_1}+\epsilon_2\frac{l}{p_2}\Big|\epsilon_1,\epsilon_2\Big),
\end{equation}
with $a$ in the Cartan subalgebra of $\mathfrak{u}(N)$. This can be rewritten (see \autoref{app.upsilon}) as
\begin{equation}
    \label{eq.4d.Z.unfact}
    Z_{\bsm p}(a;\epsilon_1,\epsilon_2)=\prod_{\alpha\in\Delta_{U(N)}}\Upsilon\Big(\ii\alpha(a)\Big|\frac{\epsilon_1}{p_1},\frac{\epsilon_2}{p_2}\Big),
\end{equation}
which is precisely the one-loop determinant of the $\mathcal{N}=2$ vector multiplet with gauge group $U(N)$ on the branched cover $\widehat{\mathbb{C}}^2_{\bsm p}$ \cite{Nishioka:2016guu}. Thus, also in four dimensions the classical and one-loop parts of the partition function agree between the refined orbifold and branched cover theories.

\paragraph{Instantons.}
In 4d there are additional, non-perturbative contributions to the partition function from (anti-)instantons. For the case of a single GW defect supported on $D_2$ and without Higgsing and twist defects, they were analysed in \cite{Kanno:2011fw}. Here, we do not provide an explicit derivation but we simply rely on the identification with the setup on the branched covers presented in \autoref{sub.branched}. Hence, we will adapt the expressions in \cite{Nishioka:2016guu} to our case of intersecting defects.

Consider the theory with GW defects \eqref{eq.defect4d} and a vev for the scalar as in \eqref{eq.higgs4d}. An instanton configuration for a subgroup $U(N)_{k,l}\subset U(N)^{p_1p_2}$ is given by  $F_{k,l}=-\star F_{k,l}$. In order to define the instanton number for this configuration we write the corresponding gauge field in terms of a smooth configuration $\bar A_{k,l}$ on $\mathbb{C}^2$:
\begin{equation}
    A_{k,l}=\bar A_{k,l}+f_1(r_1)\xi_{k,l}\dd\theta_1+f_2(r_2)\zeta_{k,l}\dd\theta_2,
\end{equation} 
where $f_i\to 0$ at $r\to\infty$ and $f_i\to1$ for $r_i\to0$. We can define the instanton number $d_{k,l}$ of $A_{k,l}$ simply as the one of $\bar A_{k,l}$, i.e.\footnote{As discussed in above, we do not consider fluxes for abelian gauge fields on the defect as the large vev of the scalar \eqref{eq.higgs4d} decouples the dynamics on the defect.} 
\begin{equation}
    \frac{1}{8\pi^2}\int_{\mathbb{C}^2-(D_1\cup D_2)}F_{k,l}\wedge F_{k,l} = d_{k,l}.
\end{equation}
Let us also define the sum of all instanton numbers
\begin{equation}\label{eq.totd}
    \sum_{k=0}^{p_1-1}\sum_{l=0}^{p_2-1}d_{k,l}=d.
\end{equation}
Fixed points under the action of $U(1)^2\times U(1)^{p_1p_2N}$ on the moduli space are isolated and labelled by a collection $\{\vec Y_{k,l}\}^{k=0,\dots,p_1-1}_{l=0,\dots,p_2-1}$ of $N$-tuples $\vec Y_{k,l}=(Y^1_{k,l},\dots,Y^N_{k,l})$ of Young diagrams. The total number of boxes for each $N$-tuple is given by $|\vec Y_{k,l}|=d_{k,l}$. 

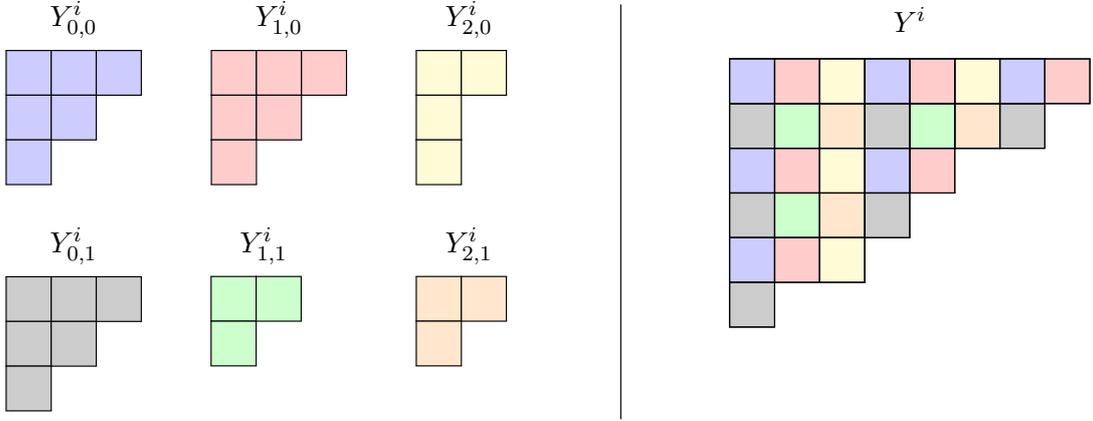
\begin{figure}[htbp]
    \centering
    \begin{tikzpicture}
        \begin{scope}[shift={(9,0)}]
        \node at (0,0) {
        \ytableaushort{\none}
     *{8,7,5,4,3,,1}
     *[*(blue!20)]{1,0,1,0,1,0}
     *[*(black!20)]{0,1,0,1,0,1}
     *[*(red!20)]{2,0,2,0,2}
     *[*(green!20)]{0,2,0,2,0}
     *[*(yellow!20)]{3,0,3,0,3}
     *[*(orange!20)]{0,3,0,3,0}
     *[*(blue!20)]{4,0,4,0}
     *[*(black!20)]{0,4,0,4}
     *[*(red!20)]{5,0,5}
     *[*(green!20)]{0,5,0}
     *[*(yellow!20)]{6,0}
     *[*(orange!20)]{0,6}
     *[*(blue!20)]{7,0}
     *[*(black!20)]{0,7}
     *[*(red!20)]{8}
        };
    
        \node at (0,2.3) {$Y^i$};
        \end{scope}
        \begin{scope}[shift={(-8,0)}]
        \node at (6,1) {
        \ytableaushort{\none}
     *[*(blue!20)]{3,2,1}
        };
        \node at (6,2.3) {$Y^i_{0,0}$};
    
        \node at (6,-2) {
        \ytableaushort{\none}
     *[*(black!20)]{3,2,1}
        };
        \node at (6,-.7) {$Y^i_{0,1}$};
    
        \node at (8.7,1) {
        \ytableaushort{\none}
     *[*(red!20)]{3,2,1}
        };
        \node at (8.7,2.3) {$Y^i_{1,0}$};
    
        \node at (8.4,-1.7) {
        \ytableaushort{\none}
     *[*(green!20)]{2,1}
        };
        \node at (8.5,-.7) {$Y^i_{1,1}$};
    
        \node at (11.1,1) {
        \ytableaushort{\none}
     *[*(yellow!20)]{2,1,1}
        };
        \node at (11.2,2.3) {$Y^i_{2,0}$};
    
        \node at (11.1,-1.7) {
        \ytableaushort{\none}
     *[*(orange!20)]{2,1}
        };
        \node at (11.2,-.7) {$Y^i_{2,1}$};
        \end{scope}
        \draw (5.2,2.5) -- (5.2,-3);
    \end{tikzpicture}
    \caption{Sewing of Young diagrams $Y^i_{k,l}$ into a single diagram $Y^i$ for $p_1=3$ and $p_2=2$ (figure adapted from \cite{Nishioka:2016guu}).}
    \label{fig.yaakov}
\end{figure}

The effect of introducing the twist defect \eqref{eq.4d.twist.explicit} is again a shift
\begin{equation}
    \ii\alpha(a)\;\rightsquigarrow\;\ii\alpha(a)+\epsilon_1\frac{k}{p_1}+\epsilon_2\frac{l}{p_2},
\end{equation} 
as was the case for the one-loop determinant. As shown in \cite{Nishioka:2016guu}, the effect of introducing the twist defect is, for fixed $i$, to sew together diagrams $Y^i_{k,l}$ for all $k,l$ into a single diagram $Y^i$, see \autoref{fig.yaakov}. In this way, we obtain a single $N$-tuple $\vec Y=(Y^1,\dots,Y^N)$, appropriate for a $U(N)$-gauge theory, and whose instanton number is \eqref{eq.totd}.

The classical part of the instanton partition function $q^{|\vec{Y}|}$ can be written as
\begin{equation}
    q^{{|\vec{Y}|}}=\prod_{k=0}^{p_1-1}\prod_{l=0}^{p_2-1}q^{|\vec{Y}_{k,l}|},
\end{equation}
since the total number of boxes does not change in the sewing procedure. The one-loop determinant for the vector multiplet involves pairs of partitions $(Y^i,Y^j)$ whose decomposition into pairs of the original diagrams follows 
\begin{equation}
    (Y^i,Y^j)_{k,l}\quad\leftrightsquigarrow\quad\{(Y^i_{k_1,l_1},Y^j_{k_2,l_2})|k_1-k_2=k\mmod p_1,\,l_1-l_2=l\mmod p_2\}.
\end{equation}
The instanton partition function for a fixed pair $(k,l)$ is given by
\begin{equation}\begin{split}
    Z^\text{Nek}_{k,l}=~&\prod_{i,j=1}^{N}\prod_{k_1,k_2=0\atop k_1-k_2=k\mmod p_1}^{p_1-1}\prod_{l_1,l_2=0\atop l_1-l_2=l\mmod p_2}^{p_2-1}\\
    &\times\!\!\prod_{s\in Y^i_{k_1,l_1}}\!\!\!\!
    \bigg(-\bigg(h_{Y^j_{k_2,l_2}}(s)+\frac{k}{ p_1}\bigg)\epsilon_1+\bigg(v_{Y^i_{k_1,l_1}}(s)+1+\frac{l}{ p_2}\bigg)\epsilon_2+\ii a_{k_2,l_2}-\ii a_{k_1,l_1}\bigg)^{-1}\\
    &\times\!\!\prod_{t\in Y^j_{k_2,l_2}}\!\!\!\! \bigg(\bigg(h_{Y^j_{k_2,l_2}}(t)+1+\frac{k}{ p_1}\bigg)\epsilon_1-\bigg(v_{Y^i_{k_1,l_1}}(t)-\frac{l}{ p_2}\bigg)\epsilon_2-\ii a_{k_2,l_2}+\ii a_{k_1,l_1}\bigg)^{-1},
\end{split}\end{equation}
where $h_{Y^i}(s)$ and $v_{Y^i}(s)$ denote the horizontal and vertical distance from the box $s=(\alpha,\beta)$ to the edge of the diagram $Y^i$. Taking the product over all factors in the refined orbifold theory we arrive at
\begin{equation}\begin{split}\label{eq.rewriting.inst}
     Z^\text{Nek}=\prod_{k=0}^{ p_1-1}\prod_{l=0}^{ p_2-1}Z_{k,l}^\text{Nek}
     =&\prod_{i,j=1}^{N}\prod_{s \in Y^{i}} \left( -h_{Y^{j}}(s)\frac{\epsilon_1}{ p_1} + (v_{Y^{i}}(s) + 1) \, \frac{\epsilon_2}{ p_2}+\ii a_{j} -\ii a_{i} \right)^{-1}\\
     &~\phantom{\prod_{i,j=1}^{N-1}}\prod_{t \in Y^{j}} \left( (h_{Y^{i}}(t) + 1)\frac{\epsilon_1}{ p_1}-v_{Y^{j}}(t) \, \frac{\epsilon_2}{ p_2}-\ii a_{j}+\ii a_{i}\right)^{-1}.
\end{split}\end{equation}
For $ p_2=1$, this reproduces the result in \cite{Nishioka:2016guu} for branched covers. These expressions can be generalised to contact instantons on $\mathbb{C}^2\times S^1$ as the free fiber is unaffected by the insertion of Gukov-Witten and twist defects. 

We conclude this section by a remark about different ways to preserve supersymmetry. For simply-connected closed four-manifolds it was shown in \cite{Pestun:2007rz,Festuccia:2018rew,Festuccia:2019akm} that rigid supersymmetric backgrounds exist where the field strength localises to instantons on some torus fixed points and anti-instantons on the others. We consider such situations in \autoref{sec.4} and obtain the anti-instanton partition function following \cite{Festuccia:2018rew} as
\begin{equation}\label{eq.exotic}
    Z^\text{anti-inst}_{\boldsymbol{ p}}(a,\bar{q},\epsilon_1,\epsilon_2)=Z^\text{inst}_{\boldsymbol{ p}}(a,\bar{q},-\epsilon_1,\epsilon_2),
\end{equation}
which holds also for complex parameters $(\epsilon_1,\epsilon_2)$.

\subsection{Adding a Free Circle}

One can extend the results for 2d and 4d theories above rather easily for the appropriate 3d theory on $\mathbb{C}_{p}\times S^1$ and 5d theory on ${\mathbb{C}}^2_{\bsm p}\times S^1$, as was already mentioned in \autoref{sub.orbi}.  
For these theories we have an additional isometric $U(1)$-action and can thus localise with respect to a $T^2$, respectively $T^3$-isometry. Because we take the $S^1$-extension to be trivial\footnote{This is motivated by the fact that, in the next section, we will glue these patches of $\mathbb{C}^r_{\bsm p}\times S^1$ into an orbifold for which the resulting $S^1$ is contractible.}, the one-loop determinant simply contains an additional product over $U(1)$-representations with equivariance parameter $\beta$ being the circumference of the $S^1$. In 3d and 5d we obtain, respectively,
\begin{align}
    &Z_p(a;\epsilon,\beta)=\prod_{\alpha\in\Delta_{U(N)}}\prod_{n_1,n_2=0}^{\infty}\Big(n_1\frac{\epsilon}{p}+n_2\beta^{-1}+\ii\alpha(a)\Big),\label{eq.3d.Z.unfact}\\[.5em]
    &Z_{\bsm p}(a;\epsilon_1,\epsilon_2,\beta)=\prod_{\alpha\in\Delta_{U(N)}}S_3\Big(\ii\alpha(a)\Big|\frac{\epsilon_1}{p_1},\frac{\epsilon_2}{p_2},\beta^{-1}\Big),\label{eq.5d.Z.unfact}
\end{align} 
where $S_3(z|\bsm\omega)=\Gamma^2(\omega/2|\bsm\omega)\Upsilon(z|\bsm\omega)$ is the triple sine function (cf. \cite{Qiu:2014oqa}). We anticipate that, when using these building blocks to compute one-loop determinants on spheres with orbifold singularities $S^{2r+1}_{\boldsymbol{\boldsymbol{p}}}$, we will need to consider fractional charges also along the trivial $S^1$ to take into account the orbifold singularities in different patches; see next section for more details.

From the observations in the previous subsections we deduce that the expressions in \eqref{eq.3d.Z.unfact}-\eqref{eq.5d.Z.unfact} are obtained also on the branched cover $\widehat{\mathbb{C}}^r_{\bsm p}\times S^1$. Assuming that this holds true also for the instanton contributions, the latter are obtained simply as the $q$-deformed 4d instanton partition function \cite{Nishioka:2016guu}.

\section{Gluing Patches}
\label{sec.4}
In the previous section we have established an equivalence between the vector multiplet on a branched cover $\widehat{\mathbb{C}}^r_{\bsm p}$ and the refined orbifold $\mathbb{C}^r_{\bsm p}$. These can be viewed as the building blocks for a theory on closed spaces, possibly with non-trivial topology. The main difficulty to obtain the partition function for such a theory is to glue the contributions from each patch in the correct way. Giving a general recipe for how this can be achieved is beyond the scope of this work\footnote{For instance, we do not discuss here how to preserve supersymmetry in the spaces with orbifold singularities which we consider. Equivalently, we do not discuss how to make the insertion of the defects supersymmetric. It is plausible that, as for the similar setup of branched covers of odd-dimensional spheres \cite{Nishioka:2013haa}, one would have to turn on a background field for the $R$-symmetry.} (and, in fact, still conjectural for smooth manifolds \cite{Bonelli:2020xps,Festuccia:2018rew}). Instead, we restrict to two kinds of spaces: spheres with orbifold singularities $S^d_{\boldsymbol{p}}$, $d=2,\dots,5$, and weighted projective spaces $\mathbb{CP}^r_{\boldsymbol{p}}$, $r=1,2$. In these cases, we give an expression for the full partition function of the refined orbifold theory in terms of building blocks, each of which containing one torus fixed point\footnote{When considering examples in 3d and 5d, these ``fixed points'' are actually fixed $S^1$-orbits. For brevity, in the general discussion we always talk about fixed points, with this subtlety understood.}. As expected, we reproduce known results on weighted projective spaces \cite{Inglese:2023wky,Martelli:2023oqk} and, confirming the equivalence between spaces with deficit and surplus angles, on branched covers of spheres \cite{Nishioka:2016guu}. We also provide predictions for entirely new results for theories on spaces with orbifold singularities.

Before we present our examples, let us comment more generally on the different aspects of the gluing procedure. The singular spaces we want to consider are closed (compact, without boundary) and simply-connected---however, possibly with non-trivial orbifold fundamental group $\pi_1^\mathrm{orb}$. We consider an open cover of this space that is adapted to equivariance, where each open subset contains a single fixed point under the torus isometry. These open subsets are diffeomorphic to $\mathbb{C}^r_{\bsm p}$ and each torus fixed point coincides with the origin of such a patch. 

Moreover, following the discussion on earlier sections, the starting point is always a gauge theory on a smooth manifold with the insertion of GW and twist defects. Hence, the gauge group needs to be chosen accordingly so that it can reproduce the singularities at all fixed points. We will explicitly show this requirement in the examples below.

\paragraph{Twisted Sectors.}
Let us first analyse fractional charges arising from the fluctuations around the BPS locus and contributing to the one-loop determinant; we will consider the Coulomb branch parameter momentarily. On each patch $\mathbb{C}^r_{\bsm p}$ the theory contains modes of fractional charges, labelling complex irreducible representations of $\pi_1^\mathrm{orb}(\mathbb{C}^r_{\bsm p})$. When gluing the partition functions of multiple patches together, the product over fractional charges might be altered, depending on the orbifold structure of the resulting space.

For instance, let us take two patches $\mathbb{C}/\mathbb{Z}_{p_1}$, $\mathbb{C}/\mathbb{Z}_{p_2}$ and glue them together so as to obtain a spindle $\mathbb{CP}^1_{(p_1,p_2)}$. Then its orbifold fundamental group\footnote{In analogy to the case of manifolds, van Kampen's theorem can be applied to compute orbifold fundamental groups.} is given by $\mathbb{Z}_{\gcd(p_1,p_2)}$. In this case, fractional modes only survive the gluing if $\gcd(p_1,p_2)>1$ and their charge is a multiple of $\frac{1}{\gcd(p_1,p_2)}$. Equivalently, this can be understood in the theory on $S^2$ with the insertion of twist defects. The boundary conditions \eqref{eq.2d.twist} need to be enforced, simultaneously, for both a $p_1$ twist defect at the north pole and a $p_2$ twist defect at the south pole. As long as $\gcd(p_1,p_2)=1$, it is impossible to satisfy both conditions. Therefore, when gluing the partition functions for the patches together, we have to restrict the product over $U(1)$-charges accordingly in each patch.

As a 3d example, consider patching two $\mathbb{C}/\mathbb{Z}_k\times S^1$, $\mathbb{C}/\mathbb{Z}_l\times S^1$ to an orbifold $S^3$. Here, the $S^1$ in one patch is identified with the generator of the orbifold fundamental group in the other patch and vice versa. In this case, when gluing the partition functions for the patches, instead of restricting to integer charges along the extra $S^1$, we need to include fractional charges corresponding to the other patch.

Unlike fractional charges, the Coulomb branch parameters in each $U(N)$-factor, being covariantly constant, are well defined also in spaces with trivial orbifold fundamental group. For example, on the spindle there are $|\boldsymbol{p}|$ such parameters which contribute to the classical piece in the partition function.

\paragraph{Regularisation.}
In \autoref{sec.3} we assumed $\epsilon_1,\epsilon_2>0$ in the infinite products appearing for the one-loop determinant (see also \autoref{app.upsilon}). However, when gluing multiple patches to a closed space it might not be possible to keep this condition for the equivariance parameters of all patches\footnote{In particular, this is not possible for the case of toric orbifolds (see \autoref{sub.4d}).}. Moreover, depending on the precise form of rigid supersymmetric background one imposes for the theory on the resulting space, the sign of either $\epsilon_1$ or $\epsilon_2$ have to be flipped in some of the patches (cf. \cite{Mauch:2021fgc}). Thus, we need to make sure that the infinite products appearing after gluing still yield a well-defined partition function. The way to do this for toric four-manifolds was explained in \cite{Festuccia:2019akm} and involves altering the domain $\mathcal{D}$ in the $\Upsilon$-function
\begin{equation}
    \Upsilon(z|a,b)=\prod_{(n,m)\in \mathcal{D}}\big(n a+m b+z\big)\big((n+1)a+(m+1)b-z\big)
\end{equation} 
from $\mathcal{D}=\mathbb{N}_0^2$ in the original definition to a different one, depending on the sign of the equivariance parameters. We expect this regularisation procedure to be unaffected by the orbifold singularities and adopt it for our examples below, producing the correct results.

\paragraph{Flux.}
When patching together spaces with non-trivial two-cycles, like the spindle $\mathbb{CP}^1_{(k,l)}$, we expect additional topological sectors to appear in the partition function, labelled by the flux through these cycles. In contrast to smooth manifolds, depending on the orbifold structure of the gauge bundle, fluxes on the orbifold are not integer-quantised. As it will be shown in explicit examples, this is precisely taken into account by the fractional charges appearing in gauge transformations \eqref{eq.quantization}. The full partition function is then obtained as a sum over all possible flux values. These appear in each patch as a shift in the Coulomb branch parameter, the precise form of which is still conjectural (even for smooth manifolds). However, extrapolating from some example computations \cite{Martelli:2023oqk,Inglese:2023wky,Mauch:2024uyt} the shift in the $\ell^\text{th}$ patch should take the general form
\begin{equation}
    \label{eq.shift.a}
    \ii\alpha(a)\;\rightsquigarrow\; \ii\alpha(a)+f(\epsilon^\ell)\frac{\alpha(\mathfrak{m})}{|\bsm p|},
\end{equation}
where $f$ is a real functions of the local equivariance parameters, collectively denoted by $\epsilon^\ell$. Flux is labelled by the element $\mathfrak{m}$ in the Cartan subalgebra of the gauge group and we assume $\alpha(\mathfrak{m})\in |\bsm p|\mathbb{Z}$ here; if this does not hold, then additional term appear in \eqref{eq.shift.a}, proportional to the remainder of $\alpha(\mathfrak{m})$ with respect to $p_i$. We show explicit expressions for these shifts in the examples below.

\subsection{Two Dimensions}
In order to see how the procedure for gluing patches works in practice, we now present an explicit example in full detail: a $U(N)$ gauge theory on a spindle\footnote{Recall that, unless specified, we are assuming $\text{gcd}( p_1, p_2)=1$.} $\mathbb{CP}^1_{\boldsymbol{p}}$, for which a known rigid supergravity background is known \cite{Inglese:2023tyc,Inglese:2023wky,Mauch:2024uyt}. The starting point to reproduce these partition functions is a $U(|\boldsymbol{p}|N)$ gauge theory on $\mathbb{CP}^1$ with the insertion of GW and twist defects at the two poles. In particular we consider a GW defect at the north pole such that $\mathbb{L}_\text{NP}=U(p_2)^{p_1}$ and a twist defect for the $\mathbb{Z}_{p_1}$ symmetry. Vice versa, at the south pole $\mathbb{L}_\text{SP}=U(p_1)^{p_2}$ and the twist defects is for the $\mathbb{Z}_{p_2}$ symmetry. Furthermore, we need to Higgs the gauge group everywhere to $U(N)^{p_1p_2}$. Note that branched covers of $\mathbb{CP}^1$ do not exist unless $p_1=p_2$, and thus a comparison is not possible as long as $\text{gcd}( p_1, p_2)=1$.

The full partition function for the refined orbifold theory for both twist and anti-twist theories is given by
\begin{equation}
    \mathcal{Z}_{\mathbb{CP}^1_{\boldsymbol{ p}}}=\sum_{\mathfrak{m}}\int \dd a\, e^{-S_\text{cl}}\cdot Z_{\mathbb{CP}^1_{\boldsymbol{ p}}}.
\end{equation}
As discussed in \autoref{sec.3}, the Coulomb branch parameter $a$ is restricted to take the same value in all $\mathfrak{u}(N)$-factors, effectively introducing a delta-function. Hence, the measure in the path integral is simply $\dd a$. The sum is over fluxes with fractional charge \cite{Inglese:2023wky}
\begin{equation}
    \frac{1}{2\pi}\int F=\frac{\mathfrak{m}}{p_1p_2}\in\frac{1}{p_1p_2}\mathbb{Z},
\end{equation}
which is exactly reproduced by the charge quantization in the refined orbifold theory \eqref{eq.quantization}. Moreover, the classical contribution is found in \cite{Inglese:2023wky,Mauch:2024uyt} and is given by
\begin{equation}\label{eq.corr}
    e^{-S_\text{cl}}=\ii\pi\ell\frac{|\boldsymbol{ p}|}{\omega_1\omega_2(\epsilon_1)}\text{Tr}\left(a^2\pm\frac{\mathfrak{m}^2}{|\boldsymbol{p}|}\right)
\end{equation}
The factor of $|\boldsymbol{ p}|$ appearing at the numerator is due to the $|\boldsymbol{ p}|$ Coulomb branch parameters contributing to the partition function on the orbifold\footnote{The factor of $|\boldsymbol{p}|$ at the numerator of \eqref{eq.corr} makes sure that, when considering an $\mathcal{N}=2$ theory on $\mathbb{CP}^1_{\boldsymbol{ p}}\times S^1$ \cite{Inglese:2023tyc}, the partition function is single-valued under large gauge transformations.}. Moreover, the $\pm$ dependence in front of the $\mathfrak{m}$ is, respectively, for twist and anti-twist theories. Finally, $\omega_1\omega_2(\epsilon)$ represents the squashing of the volume.

\paragraph{Perturbative Contributions.}
Recall that, as mentioned earlier in this section, fractional charges around the BPS locus, and entering the one-loop determinant, are not well defined globally on a spindle. Hence, we neglect them by selecting only the trivial sector of fluctuations \eqref{eq.2d.redef} at the north and south poles. The result is the following for the twist
\begin{equation}\label{eq.1looopm0.CP1top}
    Z_{\mathbb{CP}^1_{\boldsymbol{ p}}}^\text{twist}|_{\mathfrak{m}=0}=\prod_{\alpha\in\Delta}\prod_{n=0}^\infty\frac{\epsilon n-\ii\alpha( a)}{\epsilon(n+1)-\ii\alpha( a)},
\end{equation}
and for anti-twist
\begin{equation}\label{eq.1looopm0.CP1ex}
     Z_{\mathbb{CP}^1_{\boldsymbol{ p}}}^\text{anti-twist}|_{\mathfrak{m}=0}=\prod_{\alpha\in\Delta}\prod_{n=0}^\infty\frac{\epsilon n-\ii\alpha( a)}{-\epsilon(n+1)-\ii\alpha( a)}.
\end{equation}
Different regularizations for the infinite products, due to different signs in the local equivariant parameters, are taken care by placing the south pole contribution at the denominator, and by expanding for negative $n$ for the anti-twist theory at the south pole. These expressions reproduce what found in \cite{Inglese:2023wky,Mauch:2024uyt} for $\mathbb{CP}^1_{\boldsymbol{ p}}$ at $\mathfrak{m}=0$. 

Whenever $\text{gcd}( p_1, p_2)= p\neq 1$, a subsector of fractional charges $\frac{n}{ p}$ is well defined globally. Thus, our recipe instructs us to consider a subsector of the fluctuations in \eqref{eq.2d.redef} at each pole. Doing so, we find, for example for the twist theory, the following
\begin{equation}\label{eq.1looopm0.gcd}
    Z_{\mathbb{CP}^1_{\boldsymbol{ p}}}^\text{twist}|_{\mathfrak{m}=0}=\prod_{\alpha\in\Delta}\prod_{n=0}^\infty\frac{\epsilon\frac{n}{ p}-\ii\alpha( a)}{-\epsilon\left(\frac{n}{ p}+1\right)-\ii\alpha( a)}.
\end{equation}
Note that this is not the result found on $\mathbb{CP}^1_{(p,p)}$ in \cite{Hosomichi:2015pia} as, in their case, the orbifold theory only contains GW defect while twist defects are turned off.

\paragraph{One-Loop Determinant Around Fluxes.}
We now move to non-perturbative contributions, namely fluxes. As mentioned above, we compute the one-loop around fluxes dimensionally reducing the unfactorized partition function on $\widehat{S}^3_{\boldsymbol{ p}}$ along a non-trivial fiber \cite{Mauch:2024uyt}. From the reduced expression we can deduce the shifts of the Coulomb branch parameter $a^\ell$, $\ell=1,2$, at the north and south poles of $\mathbb{CP}^1_{\boldsymbol{ p}}$
\begin{equation}
    a^1=a+\ii\epsilon \frac{\alpha(\mathfrak{m})}{2 p_1 p_2}-\ii\epsilon \frac{\mmod(\alpha(\mathfrak{m}), p_1)}{ p_1},\quad a^2=a-\ii\epsilon \frac{\alpha(\mathfrak{m})}{2 p_1 p_2}+\ii\epsilon \frac{\mmod(-\alpha(\mathfrak{m}), p_2)}{ p_2}.
\end{equation}
We then find for twist
\begin{equation}\label{eq.1looopm.CP1top}
    Z_{\mathbb{CP}^1_{\boldsymbol{ p}}}^\text{twist}=\prod_{\alpha\in\Delta}\prod_{n=0}^\infty\frac{\epsilon n-\epsilon \frac{\alpha(\mathfrak{m})}{2 p_1 p_2}+\epsilon \frac{\mmod(\alpha(\mathfrak{m}), p_1)}{ p_1}-\ii\alpha( a)}{\epsilon (n+1)+\epsilon \frac{\alpha(\mathfrak{m})}{2 p_1 p_2}-\epsilon \frac{\mmod(-\alpha(\mathfrak{m}), p_2)}{ p_2}-\ii\alpha( a)},
\end{equation}
and anti-twist
\begin{equation}\label{eq.1looopm.CP1ex}
    Z_{\mathbb{CP}^1_{\boldsymbol{ p}}}^\text{anti-twist}=\prod_{\alpha\in\Delta}\prod_{n=0}^\infty\frac{\epsilon n-\epsilon \frac{\alpha(\mathfrak{m})}{2 p_1 p_2}+\epsilon \frac{\mmod(\alpha(\mathfrak{m}), p_1)}{ p_1}-\ii\alpha( a)}{-\epsilon (n+1)+\epsilon \frac{\alpha(\mathfrak{m})}{2 p_1 p_2}-\epsilon \frac{\mmod(\alpha(\mathfrak{m}), p_2)}{ p_2}-\ii\alpha( a)}.
\end{equation}
Also in this case, the numerator is the north pole contribution, the denominator the south pole contribution and the expansion in the anti-twist case is for negative $n$. This matches what is found in \cite{Inglese:2023wky,Mauch:2024uyt} for $\mathbb{CP}^1_{\boldsymbol{ p}}$.

\subsection{Four Dimensions}
\label{sub.4d}
In four dimensions we can compute, gluing building blocks \eqref{eq.4d.Z.unfact} and $\eqref{eq.rewriting.inst}$, partition functions on $S^4_{\boldsymbol{ p}}$ and $\mathbb{CP}^2_{\boldsymbol{ p}}$ \cite{Martelli:2023oqk,Mauch:2024uyt}. The former example is the first instance where fractional charges \eqref{eq.4d.redef} are well defined globally and thus the refined orbifold theory can be defined on the entire space. This will reproduce the results on branched covers $\widehat{S}^4_{(p,1)}$ \cite{Crossley:2014oea,Huang:2014pda,Nishioka:2016guu}.

\begin{figure}[htbp]
\centering
\begin{tikzpicture}
\draw[thick] (0,0) circle (3cm);
\draw[dashed] (-3,0) arc[start angle=180, end angle=360, x radius=3cm, y radius=0.7cm];
\draw (-3,0) arc[start angle=180, end angle=0, x radius=3cm, y radius=0.7cm];
\draw[green, thick] (0,-3) arc[start angle=270, end angle=90, x radius=1cm, y radius=3cm];
\draw[green, thick, dashed] (0,-3) arc[start angle=270, end angle=90, x radius=-1cm, y radius=3cm];
\draw[red, thick] (0,-3) arc[start angle=270, end angle=90, x radius=-2cm, y radius=3cm];
\draw[red, thick, dashed] (0,-3) arc[start angle=270, end angle=90, x radius=2cm, y radius=3cm];
\node[above] at (0,3) { NP};
\node[below] at (0,-3) { SP};
\node[below left] at (-1.4,-2.8) { $D_1=\;$\textcolor{green}{ $S^2_{(1)}$}};
\node[below right] at (1.4,-2.8) { $D_2=\;$\textcolor{red}{ $S^2_{(2)}$}};
\node[left] at (-2.8,1.5) {$S^4$};
\end{tikzpicture}
\caption{The four-sphere with two defects, $(\xi,\mathcal{T}_1)$ and $(\zeta,\mathcal{T}_2)$, supported on two-spheres that intersect at the poles (figure adapted from \cite{Gomis:2016ljm}).}
\label{fig.pan}
\end{figure}
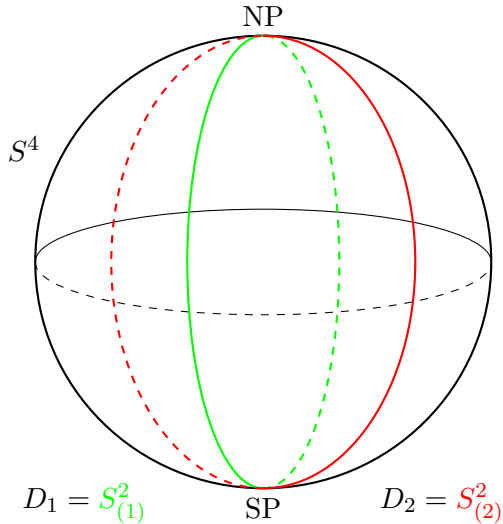

\paragraph{Four-Sphere.}
The starting point is a $U(|\boldsymbol{p}|N)$ gauge theory on $S^4$ with GW and twist defects. Therefore, we glue two refined orbifold theories at the north and south poles of an $S^4$, with the same parameter $\boldsymbol{ p}=( p_1, p_2)$. This gives our definition of a refined orbifold theory on $S^4$. The codimension two defects $\xi,\mathcal{T}_1$ and $\zeta,\mathcal{T}_2$ are placed, respectively, at $x_1=x_2=0$ and $x_3=x_4=0$ and intersect at the two poles. The support of the two defects, $D_1$ and $D_2$, are two $S^2$ as shown in \autoref{fig.pan}.
The same intersecting GW defects appear in \cite{Gomis:2016ljm}. From \autoref{fig.pan} one can also understand why fractional charges are globally defined: unlike for $\mathbb{CP}^1_{\boldsymbol{ p}}$, the singularities near both poles of $S^4_{\boldsymbol{ p}}$ are the same.

The partition function on $S^4_{\boldsymbol{ p}}$ for both topologically twisted and exotic theories is the following
\begin{equation}
    \mathcal{Z}_{S^4_{\boldsymbol{ p}}}=\int \dd a\, e^{-S_\text{cl}}\cdot Z_{S^4_{\boldsymbol{ p}}}\cdot Z^\text{inst}_{S^4_{\boldsymbol{ p}}},
\end{equation}
The classical action is
\begin{equation}
    e^{-S_\text{cl}}=e^{-\frac{8\pi^2|\boldsymbol{ p}|}{\epsilon_1\epsilon_2g_\text{YM}^2}\text{Tr}(a^2)},
\end{equation}
which matches what found in \cite{Crossley:2014oea,Huang:2014pda,Nishioka:2016guu} for the branched cover $\widehat{S}^4_{p,1}$.

The one-loop determinant on $S^4_{\boldsymbol{ p}}$ is obtained by gluing two building blocks \eqref{eq.4d.Z.unfact} for the north and south pole. For equivariant Donaldson-Witten one finds
\begin{equation}
    Z^\text{top}_{S^4_{\boldsymbol{ p}}}=\prod_{\alpha\in\Delta}\frac{\Upsilon\left(\ii a\bigg|\frac{\epsilon_1}{ p_1},\frac{\epsilon_2}{ p_2}\right)}{\Upsilon\left(\ii a\bigg|\frac{\epsilon_1}{ p_1},\frac{\epsilon_2}{ p_2}\right)}=-\alpha(a)^2,
\end{equation}
while for exotic
\begin{equation}
    Z^\text{ex}_{S^4_{\boldsymbol{ p}}}=\prod_{\alpha\in\Delta}\Upsilon\left(\ii a\bigg|\frac{\epsilon_1}{ p_1},\frac{\epsilon_2}{ p_2}\right)\Upsilon\left(-\ii a\bigg|\frac{\epsilon_1}{ p_1},\frac{\epsilon_2}{ p_2}\right).
\end{equation}
The expression for the exotic theory reproduces the result in \cite{Huang:2014pda,Crossley:2014oea,Nishioka:2016guu} on the branched cover $\widehat{S}^4_{p,1}$. A result for the exotic theory on $S^4_{p,1}$ appears in \cite{Nawata:2014nca}. However, they consider the definition of orbifold theory in \cite{Kanno:2011fw} and thus they do not include twist defects.

Finally, non-perturbative contributions arise from point-like instantons at the north and south poles \eqref{eq.rewriting.inst} 
\begin{equation}
    Z^\text{inst,top}_{S^4_{\boldsymbol{ p}}}=Z^\text{inst}_{\boldsymbol{ p}}(a,q,\epsilon_1,\epsilon_2)\cdot Z^\text{inst}_{\boldsymbol{ p}}(a,q,-\epsilon_1,\epsilon_2),
\end{equation}
\begin{equation}
    Z^\text{inst,ex}_{S^4_{\boldsymbol{ p}}}=Z^\text{inst}_{\boldsymbol{ p}}(a,q,\epsilon_1,\epsilon_2)\cdot Z^\text{anti-inst}_{\boldsymbol{ p}}(a,\bar{q},-\epsilon_1,\epsilon_2).
\end{equation}
Recall that anti-instanton partition functions have been defined in \eqref{eq.exotic}.

\paragraph{Complex Projective Space.}
The starting point in this case is a $U(|\boldsymbol{p}|N)$ gauge theory on $\mathbb{CP}^2$ where, at each of the three fixed points, we take $\mathbb{L}_i=U(Np_i)$, and we turn on twist defects for $\mathbb{Z}_{p_{i+1}}\times\mathbb{Z}_{p_{i+2}}$. To obtain the refined orbifold theory on $\mathbb{CP}^2_{\boldsymbol{p}}$ we need to Higgs the theory to $U(N)^{|\boldsymbol{p}|}$ everywhere on $\mathbb{CP}^2$. As in two dimensions, branched covers of $\mathbb{CP}^2$ do not exist for pairwise coprime $p_i$.

Fractional charges \eqref{eq.4d.redef} are not well defined in this example, thus only the trivial sector of the twisted boundary conditions \eqref{eq.4d.Tdiag} at each fixed points needs to be considered. Partition functions for exotic and topologically twisted theories on $\mathbb{CP}^2_{\boldsymbol{p}}$ have been computed in \cite{Mauch:2024uyt}, with the one-loop determinant at each flux sector for the topologically twisted theory consistent\footnote{In \cite{Martelli:2023oqk} the flux dependence is written in terms of \emph{equivariant fluxes} \cite{Bershtein:2015xfa,Bonelli:2020xps} while in \cite{Mauch:2024uyt} in terms of \emph{physical fluxes}. This obscures a direct comparison of the final expressions.} with the equivariant index on $\mathbb{CP}^2_{\boldsymbol{ p}}$ \cite{Martelli:2023oqk}. The partition functions for the two theories on the orbifold is given by
\begin{equation}
    \mathcal{Z}_{\mathbb{CP}^2_{\boldsymbol{ p}}}=\sum_{\mathfrak{m}}\int \dd a\, e^{-S_\text{cl}}\cdot Z^\text{}_{\mathbb{CP}^2_{\boldsymbol{ p}}}\cdot Z^\text{inst}_{\mathbb{CP}^2_{\boldsymbol{ p}}},
\end{equation}
The sum is over fluxes with fractional charge
\begin{equation}
    \frac{1}{2\pi}\int F=\frac{\mathfrak{m}}{|\boldsymbol{p}|}\in\frac{1}{|\boldsymbol{p}|}\mathbb{Z},
\end{equation}
which is reproduced by the charge quantization in the refined orbifold theory \eqref{eq.quantization}.
The classical contribution is given, in both cases, by
\begin{equation}
    e^{-S_\text{cl}}=e^{-\frac{8\pi^2|\boldsymbol{ p}|}{\omega_1\omega_2\omega_3(\epsilon_1,\epsilon_2)g_\text{YM}^2}\text{Tr}(a^2)}.
\end{equation}
The factor of $|\boldsymbol{p}|$ at the numerator comes from the $|\boldsymbol{p}|$ $U(N)$ factors in the refined theory. Note instead that, contrary to the $\mathbb{CP}^1_{\boldsymbol{ p}}$, the classical action has no flux-dependence. This can be achieved redefining the Coulomb branch parameter via a shift proportional to $\mathfrak{m}$ \cite{Lundin:2023tzw}. Moreover, $\omega_1\omega_2\omega_3(\epsilon_1,\epsilon_2)$ represents the squashing of the volume.

The one-loop determinants at each flux sector have been computed in \cite{Mauch:2024uyt} dimensionally reducing from $\widehat{S}^5_{\boldsymbol{ p}}$. The shifts of the Coulomb branch parameter for $\alpha(\mathfrak{m})\in p_1 p_2\mathbb{Z}$ are
\begin{equation*}\begin{split}
    \text{top:}\quad a^1=&a+\left(1-\frac{\epsilon_1+\epsilon_2}{3}\right)\frac{\alpha(\mathfrak{m})}{|\boldsymbol{ p}|},\qquad\;\,\text{ex:}\quad a^1=a+\left(\frac{1}{3}-\frac{\epsilon_1+\epsilon_2}{3}\right)\frac{\alpha(\mathfrak{m})}{|\boldsymbol{ p}|},\\
    \text{top:}\quad a^2=&a+\left(1+\frac{2\epsilon_1-\epsilon_2}{3}\right)\frac{\alpha(\mathfrak{m})}{|\boldsymbol{ p}|},\qquad\text{ex:}\quad a^2=a+\left(\frac{1}{3}+\frac{2\epsilon_1-\epsilon_2}{3}\right)\frac{\alpha(\mathfrak{m})}{|\boldsymbol{ p}|},\\
    \text{top:}\quad a^3=&a+\left(1+\frac{2\epsilon_2-\epsilon_1}{3}\right)\frac{\alpha(\mathfrak{m})}{|\boldsymbol{ p}|},\qquad\text{ex:}\quad a^3=a+\left(\frac{1}{3}+\frac{2\epsilon_2-\epsilon_1}{3}\right)\frac{\alpha(\mathfrak{m})}{|\boldsymbol{ p}|}.
\end{split}\end{equation*}
For generic values of $\alpha(\mathfrak{m})$ the shifts can be deduced from the unfactorized expression in \cite{Mauch:2024uyt} and will involve a dependence on $\mmod(\alpha(\mathfrak{m}), p_i)$.

As fractional charges are not well defined globally for pairwise $\gcd( p_\ell, p_{\ell+1})=1$, one finds that the one-loop determinant at a given flux sector for the topological theory is
\begin{equation}\begin{split}\label{eq.1looopm.CP2top}
    Z^\text{top}_{\mathbb{CP}^2_{\boldsymbol{ p}}}=&\prod_{\alpha\in\Delta}\frac{\Upsilon( \ii\alpha(a) +\beta_1^{-1}\alpha(\mathfrak{m})|\epsilon^1_1,\epsilon^1_2)\cdot\Upsilon( \ii\alpha(a) +\beta_2^{-1}\alpha(\mathfrak{m})|\epsilon^2_1,\epsilon^2_2)}{\Upsilon( \ii\alpha(a) +\beta_3^{-1}\alpha(\mathfrak{m})|\epsilon^3_1,\epsilon^3_2)}\\
    =&\prod_{\alpha\in\Delta}\Upsilon^{\tilde{\mathcal{B}}_{ \mathfrak{m},\boldsymbol{ p}}}\left(\ii\alpha(a)+\left(1-\frac{\epsilon_1+\epsilon_2}{3}\right)\frac{\alpha(\mathfrak{m})}{|\boldsymbol{ p}|}\bigg|\epsilon_1,\epsilon_2\right),
\end{split}\end{equation}
while for the exotic theory is
\begin{equation}\begin{split}\label{eq.1looopm.CP2ex}
    Z^\text{ex}_{\mathbb{CP}^2_{\boldsymbol{ p}}}=&\prod_{\alpha\in\Delta}\frac{\Upsilon( \ii\alpha(a) +\beta_1^{-1}\alpha(\mathfrak{m})|\epsilon^1_1,\epsilon^1_2)\cdot\Upsilon( \ii\alpha(a) +\beta_3^{-1}\alpha(\mathfrak{m})|\epsilon^3_1,\epsilon^3_2)}{\Upsilon( \ii\alpha(a) +\beta_2^{-1}\alpha(\mathfrak{m})|\epsilon^2_1,\epsilon^2_2)}\\
    =&\prod_{\alpha\in\Delta}\Upsilon^{\tilde{\mathcal{B}}_{ \mathfrak{m},\boldsymbol{ p}}}\left(\ii\alpha(a)+\left(\frac{1}{3}-\frac{\epsilon_1+\epsilon_2}{3}\right)\frac{\alpha(\mathfrak{m})}{|\boldsymbol{ p}|}\bigg|\epsilon_1,\epsilon_2\right).
\end{split}\end{equation}
The local equivariant parameters $\epsilon_1^\ell,\epsilon_2^\ell,\beta_\ell^{-1}$ can be found in \cite{Lundin:2021zeb}. We have also defined $\Upsilon^{\tilde{\mathcal{B}}_{ \mathfrak{m},\boldsymbol{ p}}}$-functions \cite{Lundin:2023tzw,Mauch:2024uyt}, which are generalization of $\Upsilon$-functions where the product over the integers receives contributions from generic regions of $\mathbb{Z}^2$. In this case the regions are
\begin{equation}\begin{split}
    \text{top:}\quad&\tilde{\mathcal{B}}_{\mathfrak{m},\boldsymbol{ p}}=\{(\tilde{n}_1,\tilde{n}_2)\in\mathbb{Z}^2_{\geq 0}\; |\; \tilde{n}_1+\tilde{n}_2\leq \frac{\alpha(\mathfrak{m})}{\boldsymbol{p}}\},\\
    \text{ex:}\quad&\tilde{\mathcal{B}}_{\mathfrak{m},\boldsymbol{ p}}=\{(\tilde{n}_1,\tilde{n}_2)\in\mathbb{Z}^2_{\geq 0}\; |\; \tilde{n}_1+\tilde{n}_2\geq \frac{\alpha(\mathfrak{m})}{\boldsymbol{p}}\}.
\end{split}\end{equation}
Finally, instanton contributions from the three fixed points of $\mathbb{CP}^2_{\boldsymbol{p}}$ are the following
\begin{equation}
    Z^\text{inst,top}_{\mathbb{CP}^2_{\boldsymbol{ p}}}=\prod_{\ell=1}^3 Z^\text{inst}_{\boldsymbol{p}}(a^\ell,\mathfrak{m},q,\epsilon^\ell_1,\epsilon^\ell_2),
\end{equation}
\begin{equation}
    Z^\text{inst,ex}_{\mathbb{CP}^2_{\boldsymbol{ p}}}=Z^\text{inst}_{\boldsymbol{ p}}(a^1,\mathfrak{m},q,\epsilon_1,\epsilon_2)\cdot\prod_{\ell=2}^3 Z^\text{anti-inst}_{\boldsymbol{ p}}(a^\ell,\mathfrak{m},\bar{q},\epsilon^\ell_1,\epsilon^\ell_2).
\end{equation}
Whenever $p=\gcd(p_1,p_2)=\gcd(p_2,p_3)=\gcd(p_1,p_3)$, fractional charges are well defined globally and need to be considered in the partition function.

\subsection{Odd Dimensions}
We compute partition functions on $S^{2r+1}_{\boldsymbol{ p}}$ gluing perturbative contributions \eqref{eq.3d.Z.unfact}-\eqref{eq.5d.Z.unfact} and, in 5d, also Nekrasov partition functions \eqref{eq.rewriting.inst} at each fixed fiber. These results will reproduce earlier computations for partition functions on branched covers of odd-dimensional spheres \cite{Nishioka:2013haa,Nishioka:2016guu}. Following the discussion above, in these spaces fractional charges are well defined globally. Expressions for the full partition function on branched covers can be found in \cite{Nishioka:2016guu}, here we focus on the perturbative parts only. Both in 3d and 5d, the starting point is a $U(|\boldsymbol{p}|N)$ gauge theory on $S^{2r+1}$ with a choice of GW and twist defect which reproduces the singularity structure.

Before presenting the explicit results, let us make a general comment which apply to both 3d and 5d. The partition functions we find match those on branched covers of $S^3$ and $S^5$ \cite{Nishioka:2013haa,Nishioka:2016guu}. These are equivalently computed on the resolved spaces: squashed spheres with squashing parameters determined by the singularities. Upon dimensional reduction along a locally free $S^1$-action, these partition functions encode the full partition function, at all flux sectors, on the spindle \cite{Mauch:2024uyt}. Hence, the partition function on the spindle is completely determined by that on a higher dimensional smooth manifold. This property is reminiscent of a similar behaviour in gravity, where 4d black holes with a near horizon with a spindle factor, uplift to completely smooth 10d/11d solutions \cite{Ferrero:2020twa}. It would be worthwhile to further explore this relation.

\paragraph{Three Sphere.}
To derive the perturbative partition function on $S^3_{\boldsymbol{ p}}$ we glue two building blocks \eqref{eq.3d.Z.unfact} for a $\boldsymbol{p}$-refined orbifold theory on $\mathbb{C}\times S^1$ at the north pole fixed fiber and at the south pole fixed fiber. Thus, we find
\begin{equation}\label{eq.S31loop}
    Z_{S^3_{\boldsymbol{ p}}}=\prod_{\alpha\in\Delta}\frac{\prod_{n_1,n_2=0}^\infty\left(\frac{\omega_1}{ p_1}n_2+\frac{\omega_2}{ p_2}n_2+\ii\alpha(a)\right)}{\prod_{n_1,n_2=1}^\infty\left(\frac{\omega_1}{ p_1}n_2+\frac{\omega_2}{ p_2}n_2-\ii\alpha(a)\right)},
\end{equation}
where numerator/denominator represent north/south pole contributions. For branched covers this reproduces the result in \cite{Nishioka:2013haa,Nishioka:2016guu}. As a non-trivial test of our results, one can check that dimensionally reducing along the ``orbifold'' Hopf fiber of $S^3_{\boldsymbol{p}}$, along the lines of \cite{Mauch:2024uyt}, it is possible to recover the one-loop determinant at all flux sectors for both the twist and anti-twist theories on $\mathbb{CP}^1_{\boldsymbol{p}}$ in \eqref{eq.1looopm.CP1top}-\eqref{eq.1looopm.CP1ex}.

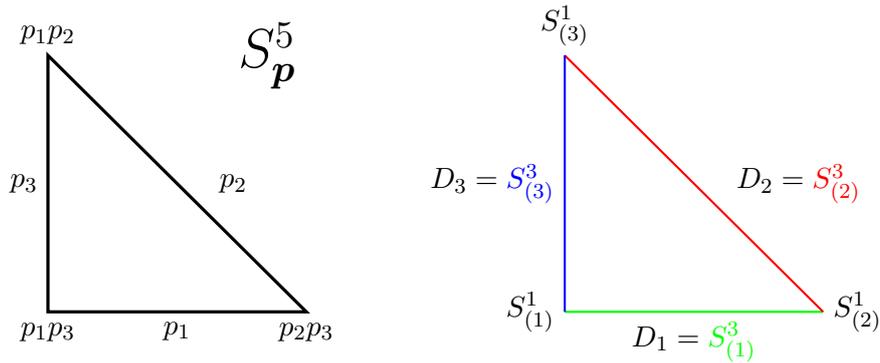
\begin{figure}
\centering
\begin{tikzpicture}[scale=.85]
\draw[line width=1.2pt] (0,0) node[below]{$ p_1 p_3$}--(4,0) node[below]{$ p_2 p_3$}--(0,4) node[above]{$ p_1 p_2$}--cycle;
\node at (2,0) [below,align=center]{$ p_1$};
\node at (0,2) [left,align=center]{$ p_3$};
\node at (2.5,2) [right,align=center]{$ p_2$};
\node[left] at (4,4) {\huge $S^5_{\boldsymbol{ p}}$};
\begin{scope}[shift={(8,0)}]
    \draw[green, thick] (0,0)--(4,0);
    \draw[red, thick] (4,0)--(0,4);
    \draw[blue, thick] (0,4)--(0,0);
    \node at (2,0) [below,align=center]{$D_1=\;$\textcolor{green}{$S^3_{(1)}$}};
    \node at (2.5,2) [right,align=center]{$D_2=\;$\textcolor{red}{$S^3_{(2)}$}};
    \node at (0,2) [left,align=center]{$D_3=\;$\textcolor{blue}{$S^3_{(3)}$}};
    \node at (0,0) [left]{$S^1_{(1)}$};
    \node at (4,0) [right]{$S^1_{(2)}$};
    \node at (0,4) [above]{$S^1_{(3)}$};
\end{scope}
\end{tikzpicture}
\caption{Left-hand side: singularity structure of $S^5_{\boldsymbol{ p}}$ and $\widehat{S}^5_{\boldsymbol{p}}$. The $ p_i$ represent either orbifold singularities or branching indices. Right-hand side: the refined orbifold theory on $S^5$ with the insertion of Gukov-Witten defects $\xi,\mathcal{T}_1$ on $D_1$, $\zeta,\mathcal{T}_2$ on $D_2$ and $\chi,\mathcal{T}_3$ on $D_3$. The intersection of each defect is along an $S^1$, for example $D_1\cap D_3=S^1_{(1)}$.}
\label{fig.triangles}
\end{figure}

\paragraph{Five Sphere.}
We consider spaces $S^5_{\boldsymbol{ p}}$, which have three fixed fibers characterized by different $T^2\subset T^3$ collapsing, where $T^3$ is the Cartan of the isometry group. To compute the partition function we glue, at each of the three fixed fibers, a $\boldsymbol{p}$-refined orbifold theory. All these refined orbifold theories are taken on $\mathbb{C}^2\times S^1$. The support of each pair of defects is an $S^3$ connecting two fixed fibers, as show in \autoref{fig.triangles}. The defects intersect along an $S^1$ at the three fixed fibers. This construction defines the refined orbifold theory on $S^5$.

To compute the perturbative partition function we glue one contribution \eqref{eq.5d.Z.unfact} for each of the three fixed fibers of $S^5_{\boldsymbol{ p}}$ and find\footnote{The relation between equivariant parameters $\epsilon^\ell,\beta^\ell$ and squashing parameters $\omega_i$ can be found in \cite{Lundin:2021zeb}.}
\begin{equation}\begin{split}\label{eq.1looop.S5}
    Z_{S^5_{\boldsymbol{ p}}}&=\prod_{\alpha\in\Delta}\frac{S_3\left( \ii\alpha(a) \bigg|\frac{\epsilon_1^1}{ p_1},\frac{\epsilon_2^1}{ p_2},\frac{\beta_1^{-1}}{ p_3}\right)\cdot S_3\left( \ii\alpha(a)\bigg|\frac{\epsilon_1^2}{ p_2},\frac{\epsilon_2^2}{ p_3},+\frac{\beta_2^{-1}}{ p_1}\right)}{S_3\left( \ii\alpha(a)\bigg|\frac{\epsilon_1^3}{ p_3},\frac{\epsilon_2^3}{ p_1},\frac{\beta_3^{-1}}{ p_2}\right)}\\
    &=\prod_{\alpha\in\Delta}S_3\left( \ii\alpha(a)\bigg|\frac{\omega_1}{ p_1},\frac{\omega_2}{ p_2},\frac{\omega_3}{ p_3}\right),
\end{split}\end{equation}
which reproduces \cite{Alday:2014fsa,Hama:2014iea,Nishioka:2016guu} for $\widehat{S}^5_{ p,1,1}$. As for $S^3_{\boldsymbol{p}}$, also on $S^5_{\boldsymbol{p}}$ we can dimensionally reduce and reproduce the results for both topologically twisted and exotic theories on $\mathbb{CP}^2_{\boldsymbol{p}}$ \cite{Mauch:2024uyt}.

\section{Discussion}\label{sec.6}
In this paper we introduced what we call a $U(N)$ refined orbifold theory for the 2d $\mathcal{N}=(2,2)$ vector multiplet on $\mathbb{C}/\mathbb{Z}_p$ and the 4d $\mathcal{N}=2$ vector multiplet on $\mathbb{C}/\mathbb{Z}_{p_1}\times\mathbb{C}/\mathbb{Z}_{p_2}$, using Gukov-Witten defects and twist defects. We pointed out the similarities of its formulation with that for theories on branched covers $\widehat{\mathbb{C}}_p,\widehat{\mathbb{C}}^2_{\bsm p}$, and argued for an equivalence between the two. In particular, we showed that their partition functions agree. Adding a free fiber we also computed partition functions on $\mathbb{C}/\mathbb{Z}_{p}\times S^1$ and $\mathbb{C}/\mathbb{Z}_{p_1}\times\mathbb{C}/\mathbb{Z}_{p_2}\times S^1$. By gluing these building blocks carefully, we gave expressions for partition functions on orbifolds of spheres $S^d_{\boldsymbol{p}}$ and weighted projective spaces. These reproduce results in the literature. 

In fact, the insertion of Gukov-Witten defects in 4d is reminiscent of another property of orbifold singularities. Gukov-Witten defects can be seen as probe cosmic strings \cite{Kibble:1976sj,VILENKIN1985263,Heidenreich:2021xpr}, in the same way as 't Hooft lines are interpreted as very heavy (non-dynamical) monopoles. When a cosmic string is created, possibly during a spontaneous symmetry breaking process, its mass affects spacetime. Via a sort of gravitational Aharonov-Bohm effect, it creates an orbifold singularity transversal to its codimension-two worldvolume. Reinterpreting our refined orbifold theory in this language, we find that to describe SQFTs on orbifolds one has to consider the insertion of $ p$ different probe cosmic strings. Such insertion gives rise to a $\mathbb{Z}_{\boldsymbol{p}}$ global symmetry for which we can turn on the twist defects.

\paragraph{Future Directions.}
\begin{itemize}
    \item We do not currently know how to extend the refined orbifold to fundamental matter in a way consistent with results on branched covers. In this case, also the GW defect has a non-trivial effect even after breaking the gauge group to $\mathbb{L}$. A possibility would be to Higgs the theory with GW defect by a vev that is rotated away from the diagonal by a transformation that diagonalises \eqref{eq.T2d} (resp. \eqref{eq.4d.twist.explicit} in 4d). Without matter, this non-diagonal Higgsing simply replaces the effect of the twist defect.
    \item We argued that the $\mathbb{Z}_{\boldsymbol{p}}$ global symmetry is present in the full theory with GW defects, before turning on a large vev for the scalar. It would be interesting to study the insertion of twist defects in this setup and study the combined effect of the two defects. For example, one would have to take into account degrees of freedom on the codimension two (or four) insertion locus of the defects or, on the orbifold side, on the singularity locus.
    \item As mentioned in the main body of the paper, the building block employed in \cite{Inglese:2023wky} to compute the spindle index is the equivariant index on orbifolds \cite{10.1215/S0012-7094-96-08226-5}. While the final results do agree with ours, the expression for the building block is different. In particular, in \cite{Inglese:2023wky}, in the sum over $ p$ contributions, large cancellations occur. It would be interesting to compare the two approaches.
\end{itemize}

\paragraph*{Acknowledgments}
        We are grateful to Guido Festuccia, Dario Martelli and Itamar Yaakov for stimulating discussions and correspondence on the subject. RM acknowledges support from the Centre for Interdisciplinary Mathematics at Uppsala University. LR acknowledges support from the Shuimu Tsinghua Scholar Program.

\appendix
\section{$\Upsilon$-Functions}
\label{app.upsilon}

In this appendix we introduce $\Upsilon$-functions, which are the main ingredient of one-loop determinants, and give some useful identities (see \cite{Zamolodchikov:1995aa,Hama:2012bg,Nishioka:2016guu}). In particular, one of them proves the equivalence between \eqref{eq.4d.Z.fact} and \eqref{eq.4d.Z.unfact}. 

Let $\bsm\epsilon=(\epsilon_1,\dots,\epsilon_r)$ with $\epsilon_i\ge0$ and $z,s\in\mathbb{C}$.

\begin{defi}
    (Barnes zeta function) The Barnes zeta function is defined as
    \begin{equation}
        \zeta(s,z|\bsm\epsilon)=\sum_{n_1,\dots,n_r\ge0}(n_1\epsilon_1+\dots+n_r\epsilon_r+z)^{-s}.
    \end{equation}
\end{defi}
Note that for $r=1,\epsilon_1=1$ and $z=1$ this is the Riemann zeta function. The Barnes zeta function can also be expressed in an integral representation:
\begin{equation}
    \zeta(s,z|\bsm\epsilon)=\frac{1}{\Gamma(s)}\int_0^\infty\frac{\e{-zt}}{\prod_{i=1}^{r}(1-\e{-\epsilon_i t})}t^{s-1}\dd t.
\end{equation}
In this representation it is easy to see that the following identity about factorisation holds:
\begin{equation}
    \label{eq.app.Barnes}
    \forall p\in \mathbb{N}, i=1,\dots,r:\;\zeta\Big(s,z\Big|\epsilon_1,\dots,\frac{\epsilon_i}{p},\dots,\epsilon_r\Big)=\sum_{k=0}^{p-1}\zeta\Big(s,z+\frac{k\epsilon_i}{p}\Big|\bsm\epsilon\Big).
\end{equation} 
\begin{defi}
    (Multiple gamma function) The multiple gamma function $\Gamma(z|\bsm\epsilon)$ is defined as
    \begin{equation}
        \Gamma(z|\bsm\epsilon)=\exp\left(\frac{\partial}{\partial s}\zeta(s,z|\bsm\epsilon)|_{s=0}\right).
    \end{equation}
\end{defi}
\begin{defi}
    ($\Upsilon$-function) Notate $\epsilon=\epsilon_1+\dots+\epsilon_r$. Then the $\Upsilon$-function is defined in terms of multiple gamma functions as
    \begin{equation}
        \Upsilon(z|\bsm\epsilon)=\frac{\Gamma^2(\epsilon/2|\bsm\epsilon)}{\Gamma(z|\bsm\epsilon)\Gamma(\epsilon-z|\bsm\epsilon)}.
    \end{equation}
\end{defi}
From this definition one can derive various identities for $\Upsilon(z|\bsm\epsilon)$. Most notably, using the definitions above and \eqref{eq.app.Barnes}, one obtains the following factorisation property:
\begin{equation}
    \Upsilon\Big(z\Big|\epsilon_1,\dots,\frac{\epsilon_i}{p},\dots,\epsilon_r\Big)=\prod_{k=0}^{p-1}\Upsilon\Big(z+\frac{k\epsilon_i}{p}\Big|\bsm\epsilon\Big).
\end{equation}

\bibliographystyle{utphys}
\bibliography{main}

\end{document}